\documentclass[10pt]{article}
\usepackage{wrapfig,array}
\usepackage{epsfig,cancel,amsthm,amssymb, todonotes}
\usepackage{color,tikz}
\usetikzlibrary{decorations.markings,arrows, calc}
\usepackage{color}
\usepackage{graphicx,framed,verbatim,caption,amsbsy}
\usepackage[colorlinks=true, pdfstartview=FitV, linkcolor=darkblue, citecolor=darkblue, urlcolor=darkblue]{hyperref}
\definecolor{shadecolor}{rgb}{0.9, 0.9, 0.86}
\definecolor{darkgreen}{rgb}{0.2, 0.5,  0}
\definecolor{darkblue}{rgb}{0.1,0.1,0.45}

\def\&{\vspace{-5pt}&}

\def\cA{{\mathcal A}}

\def \p{\mathbf p}
\def \q{\mathbf q}

\def \eqref#1{(\ref{#1})}
\def \& {&\hspace{-10pt}}

\def \wt{\widetilde}

\renewcommand{\d}{\mathrm d}
\newcommand{\pa}{\partial}       
\newtheorem{theorem}{Theorem}[section]
\newtheorem{example}[theorem]{Example}
\newtheorem{exercise}[theorem]{Exercise}

\newtheorem{lemma}[theorem]{Lemma}
\newtheorem{remark}[theorem]{Remark}

\newtheorem{proposition}[theorem]{Proposition} 
\newtheorem{corollary}[theorem]{Corollary} 
 
\newtheorem{definition}[theorem]{Definition}

\def\le{\left}
\def\ri{\right}
\def\ds{\displaystyle}

\def\bt{\begin{theorem}}
\def\et{\end{theorem}}
\def\bc{\begin{corollary}}
\def\ec{\end{corollary}}
\def\bx{\begin{example}}
\def\ex{\end{example}}
\def\bxr{\begin{exercise}\small}
\def\exr{\end{exercise}}
\def\bl{\begin{lemma}}
\def\el{\end{lemma}}
\def\bd{\begin{definition}}
\def\ed{\end{definition}}
\def\bp{\begin{proposition}}
\def\ep{\end{proposition}}

\def\br{\begin{remark}}
\def\er{\end{remark}}

\def\be{\begin{eqnarray}}
\def\ee{\end{eqnarray}}
\def \Tr {\mathrm{Tr}\,}
\def\&{\hspace{-15pt}&}
\def\bea{\begin{eqnarray}}
\def\eea{\end{eqnarray}}
\def\beas{\begin{eqnarray*}}
\def\eeas{\end{eqnarray*}}
\def\B{\boldsymbol {\mathfrak B}}
\def \pa{\partial}
\def\C{{\mathbb C}}

\def\R{{\mathbb R}}
\def\N{{\mathbb N}}

\def\wh{\widehat}

\def\Z{{\mathbb Z}}

\def\a{\alpha}
\def\K{\mathcal K}

\def\1{{\bf 1}}
\def\s{ {\sigma}} 

\def\diag{{\mathrm{diag}}}
\def\Id{{\mathrm{Id}}}
\def\ot{{\otimes}}

\def \A{\mathbf A}
\def \B{\mathbf B}
\def \C{\mathbf C}
\def\a{\alpha}
\def \diag { {\mathrm {diag}}}

\def \b{\beta}
\def \ad{\mathrm {ad}}
\def\form{\mathrm{form}}

\def\QED {\hfill $\blacksquare$\par\vskip 3pt}

\makeatletter
\@addtoreset{equation}{section}
\makeatother

\begin{document}

\baselineskip 14pt plus 1pt minus 1pt

\begin{flushright}
\end{flushright}
\vspace{0.2cm}
\begin{center}
\begin{Large}
\textbf{Noncommutative Painlev\'e\ equations and  systems of Calogero type} 
\end{Large}
\end{center}
\bigskip
\begin{center}
M. Bertola$^{\dagger\ddagger \clubsuit}$\footnote{Marco.Bertola@\{concordia.ca, sissa.it\}},  
M. Cafasso $^{\diamondsuit}$ \footnote{cafasso@math.univ-angers.fr},
V. Rubtsov $^{\diamondsuit}$ \footnote{Vladimir.Roubtsov@univ-angers.fr}
\\
\bigskip
\begin{minipage}{0.7\textwidth}
\begin{small}
\begin{enumerate}
\item [${\dagger}$] {\it  Department of Mathematics and
Statistics, Concordia University\\ 1455 de Maisonneuve W., Montr\'eal, Qu\'ebec,
Canada H3G 1M8} 
\item[${\ddagger}$] {\it SISSA/ISAS, via Bonomea 265, Trieste, Italy }
\item[${\clubsuit}$] {\it Centre de recherches math\'ematiques,
Universit\'e de Montr\'eal\\ C.~P.~6128, succ. centre ville, Montr\'eal,
Qu\'ebec, Canada H3C 3J7} 
\item [${\diamondsuit}$] {\it LAREMA, Universit\'e d'Angers\\ 2 Boulevard Lavoisier, 49045 Angers, France.}
\end{enumerate}
\end{small}
\end{minipage}
\vspace{0.5cm}
\end{center}
\bigskip
\begin{center}
\begin{abstract}
All Painlev\'e equations can be written as  a time--dependent Hamiltonian system, and as such they admit a natural generalization to the case of several particles with an interaction of Calogero type (rational, trigonometric or elliptic). Recently, these systems of interacting particles have been proved to be relevant in the study of $\beta$--models. An almost two decade old open question  by Takasaki asks whether  these multi-particle systems can be understood as isomonodromic equations, thus extending the Painlev\'e correspondence. In this paper we answer in the affirmative by displaying explicitly suitable  isomonodromic  Lax pair formulations. As an application of the isomonodromic representation we provide a construction  based on  discrete Schlesinger transforms, to produce solutions for these systems for special values of the coupling constants,  starting from uncoupled ones; the method is  illustrated for the case of the second Painlev\'e equation.
\end{abstract}
\end{center}
\tableofcontents

\section{Introduction}

The celebrated Painlev\'e--Calogero correspondence \cite{ManinPainleve,TakasakiPaCa} consists in the remarkable observation that all Painlev\'e equations can be written in the particular form
$$\ddot{q} = -V(q;t)$$
for some function $V$ depending explicitly on both the dependent and independent variables.
In other words, they all can be thought of as systems with a physical  Hamiltonian
$$H(p,q;t) := \frac{p^2}2 + V(q;t)$$
(and standard symplectic form), describing the motion of a particle in a time--dependent potential.

For the  sixth Painlev\'e equation (PVI), this result is due to Manin \cite{ManinPainleve}  (an expression of the sixth Painlev\'e equation in terms of elliptic functions is already present in the original work of Painlev\'e, \cite{Painleve})
  and the Hamiltonian is written as
\be\label{HamPVI}
	H = \frac{p^2}2 -  \sum_{i = 0}^3 g_i\wp( q + \omega_i),
\ee
where $\wp$ is the Weierstrass function and the four parameters $\{g_i,\, i = 0,\ldots,3\}$ are in correspondence with the parameters in the PVI equation (see the subsection \ref{subsPVI} for their explicit expression). The four coefficients $\omega_i$ reads
$$(\omega_0,\omega_1,\omega_2,\omega_3) = (0,1/2,-(1+\tau)/2,\tau/2),$$
where $\tau$ is the modular parameter and plays the role of the independent  time variable $t$.\\

It was Levin and Olshanetsky \cite{LeOl} that pointed out that the Hamiltonian \eqref{HamPVI} corresponds to the rank--one case of Inozemtsev's extensions \cite{Ino} of the Calogero-Moser systems, when one considers $\tau$ as a parameter and not as the independent variable. Takasaki, in \cite{TakasakiPaCa}, found the Calogero form of each of the Painlev\'e equations by observing that they all can be deduced from the Okamoto's ones \cite{OkaHam} by some explicit canonical transformation. Moreover, he extended the Calogero--Painlev\'e correspondence to what he called ``multi--component'' Painlev\'e equations. More precisely, he proved that there exist some canonical transformations between: 
\begin{itemize}
	\item The Inozemstev's {\it multi--component} systems and their degenerations (from PVI to PI).
	\item A multi--component generalization of the Okamoto polynomials.
\end{itemize}
For the reader's convenience, we report here the list of the Hamiltonians of what we call ``Calogero-Painlev\'e'' systems.
\bea
	\tilde H_{VI}:&&  \sum_{j = 1}^n \Bigg(\frac{p_j^2}2 + \sum_{\ell = 0}^3 g_\ell^2 \wp(q_j + \omega_\ell)\Bigg) + g_4^2 \sum_{j \neq k} \Bigg(\wp(q_j - q_k) + \wp(q_j + q_k) \Bigg).
	\nonumber\\
	\tilde H_V: && \sum_{j = 1}^n \Bigg(\frac{p_j^2}2 - \frac{\alpha}{\sinh^2(q_j/2)} - \frac{\beta}{\cosh^2(q_j/2)} + \frac{\gamma t}2\cosh(q_j) + \frac{\delta t^2}8\cosh(2 q_j) \Bigg) +
	\nonumber \\
	&& \hspace{0.4cm} + g_4^2 \sum_{j \neq k}\Bigg( \frac{1}{\sinh^2((q_j - q_k)/2)} + \frac{1}{\sinh^2((q_j + q_k)/2)}\Bigg).
	\nonumber \\
	\tilde H_{IV}: && \sum_{j = 1}^n \Bigg(\frac{p_j^2}2 - \frac{1}2 \left(\frac{q_j}{2}\right)^6 - 2t \left(\frac{q_j}2\right)^4 -2 (t^2 - \alpha) \left(\frac{q_j}{2}\right)^2 + \beta \left(\frac{q_j}2\right)^{-2} \Bigg) +g_4^2 \sum_{j \neq k} \Bigg(\frac{1}{(q_j - q_k)^2} + \frac{1}{(q_j + q_k)^2}\Bigg).
	\nonumber\\
	\tilde H_{III}:&&  \sum_{j = 1}^n \Bigg(\frac{p_j^2}2 - \frac{\alpha}4{\rm e}^{q_j} + \frac{\beta t}4{\rm e}^{-q_j} - \frac{\gamma}8{\rm e}^{2 q_j} + \frac{\delta t^2}8{\rm e}^{-2q_j} \Bigg) + g_4^2 \sum_{j \neq k} \frac{1}{\sinh^2\big((q_j - q_k)/2\big)}.
\nonumber  \\
	\tilde H_{II}: && \sum_{j = 1}^n \Bigg(\frac{p_j^2}2 - \frac{1}2\Big(q_j^2 + \frac{t}2 \Big)^2 - \alpha q_j \Bigg) + g_4^2 \sum_{j \neq k}\frac{1}{(q_j - q_k)^2}.
	\nonumber \\
	\tilde H_I: && \sum_{j = 1}^n \Bigg(\frac{p_j^2}2 - 2q_j^3 - tq_j \Bigg) + g_4^2\sum_{j \neq k}\frac{1}{(q_j - q_k)^2}.
	\label{takalista}	
\ee

In the concluding remarks of his paper, Takasaki stated that ``{\it a central issue will be to find an isomonodromic description of the multi--component Painlev\'e equations. If such an isomonodromic description does exist, it should be related to a new geometric structure}''. The main result of this paper is exactly the description of this isomonodromic formulation:
\begin{theorem}
	All the Hamiltonian systems in the Takasaki list \eqref{takalista} have an isomonodromic formulation in terms of a $2n\times 2n$ Lax pair, where $n$ is the number of particles.
\end{theorem}
In each case the dynamical variables appear as the {\it eigenvalues} of an $n\times n$ matrix, which we denote hereafter $\q$,  evolving in accordance to a matrix (i.e. {\it non-commutative}) version of the corresponding Painlev\'e\ equation. 
Our result is constructive and the Lax pairs are explicitly written. They are obtained by Hamiltonian reduction \`a la Kazhdan-Konstant-Sternberg \cite{KKS} on the Lax systems for the matrix Painlev\'e equations recently written by Kawakami \cite{KKS} except for Painlev\'e\ II, where the Lax pair is closer to the standard Flaschka--Newell Lax pair \cite{FN}.
 These are particular examples of the so--called simply laced isomonodromy systems introduced by P. Boalch in \cite{Boalch}. More precisely, they correspond to hyperbolic Dinkyn diagrams obtained by adding one leg to the affine diagrams associated to the corresponding ``scalar'' Painlev\'e equations.

An implicit result of the isomonodromic representation is the remarkable property (which is subsumed by the naming convention in the literature  but was never proved), that all the equations satisfy the Painlev\'e\ property, namely, the solutions $\q(t)$ have only movable poles when considered as functions of the complex time $t$ (note, however, that the {\it eigenvalues} of $\q(t)$ in general are not meromorphic functions, only their symmetric polynomials are).\\

A second important observation is that the space of initial data of each equation is identifiable with a suitable  manifold of (generalized) monodromy data, which is an algebraic variety that can be easily written in explicit form. We do so here only in the case of the second Painlev\'e\ system  (Section \ref{PIIsection}, and particularly Theorem \ref{thmqstokes}) but the construction is clearly general, with the due modifications. This manifold plays, in this setting, a role similar to the one played by the completed Calogero--Moser space  $\mathcal C_n$ (the {\it adelic Grassmannian} of \cite{WilsonAdelic}) in the ``classical" setting (see Remark \ref{StokesetWilson}).
\\

The structure of the paper is as follows: in the Section \ref{sectiongeneralscheme} we explain the general structure of the construction, while in Section \ref{sectionlist} we provide details for each of the equations. The Section \ref{PIIsection} focuses on the case of the second Painlev\'e equation. Here we start studying its Stokes phenomena in a more general setting (instead of a matrix-valued Painlev\'e II equation we study the equation with values in an arbitrary non--commutative algebra, as in \cite{RetakhRubtsov}) and we relate it to the quantization of the monodromy manifold of the Flashka--Newell Lax pair for the second Painlev\'e equation \cite{MazRou}. Finally, in the subsection \ref{specsols}, we show how to use discrete Schlesinger transformations to construct solutions of the second Calogero--Painlev\'e system, out of $n$ ``decoupled'' solution of the second Painlev\'e equation.

\subsection{Quantization, $\beta$--models and open questions}

Zabrodin and Zotov, in \cite{QPC1ZZ,QPC2ZZ}, provided a quantized version of the Calogero--Painlev\'e correspondence. Namely, they proved that for each of the Painlev\'e equation it exists a Lax pair such that the first component $\psi(z;t)$ of its eigenfunction satisfies the equation\footnote{Up to a shift on the parameters of the equation.}
	\be\label{ZabrodinZotov}
		\partial_t \psi(z;t) = \Big( \tilde H(z,\partial_z) - \tilde H(q,p) \Big)\psi(z;t),
	\ee
where $\tilde H$ here indicates any of the Hamiltonian in the Takasaki's list, in the case $n = 1$. It would be interesting to extend this quantum Calogero--Painlev\'e correspondence to the higher rank cases $n > 1$.\\ 

The quantum Painlev\'e equations have interesting applications in the theory of $\beta$--models, which are statistical models generalising (unitary-invariant) random matrices (which correspond to $\beta = 2$). For instance, as the fluctuations  of the largest eigenvalue of a random matrix (under suitable assumptions) is governed by the Hasting Mc--Leod solution of the second Painlev\'e equation \cite{TracyWidomLevel}, in the same way the position of the largest particle in a $\beta$--ensemble is governed by the $\beta$--dependent quantum Painlev\'e II equation
\be\label{qPII}
	\Bigg(\frac{\beta}2 \frac{\partial}{\partial s} + \frac{\partial^2}{\partial \xi^2} + (s - \xi^2)\frac{\partial}{\partial \xi} \Bigg)\mathcal F(\xi;s) = 0,
\ee 
as discovered in \cite{BloemViragI}. The equation \eqref{qPII}, indeed, for $\beta = 2$, is of the same type of \eqref{ZabrodinZotov}, except for the fact that a different Hamiltonian 
$$H := \frac{p^2}2 - (q^2 + t)p$$ 
had been used, and that $F(\xi;s) = {\rm e}^{K}\psi(\xi;s)$,  where $K$ is the (indefinite) integral of the Hamiltonian.
In \cite{RumanovqPII}, Rumanov provided an important step in extending the quantum Calogero-Painlev\'e correspondence to general values of $\beta$. Namely, consider an arbitrary Lax pair  

\bea\label{RumanovLaxPair}
\le(\pa_\xi 
-\le[
\begin{array}{cc}
L_1(\xi;s) & L_+(\xi;s)\\
L_-(\xi;s) & L_2(\xi;s)
\end{array}\ri]\ri)
\le[\begin{array}{c}
\mathcal F(\xi;s)
\\
\mathcal G(\xi;s)
\end{array}\ri] &=& 0, \nonumber
\\
\\ 
\le(\pa_s
-\le[
\begin{array}{cc}
B_1(\xi;s) & B_+(\xi;s)\\
B_-(\xi;s) & B_2(\xi;s)
\end{array}\ri]\ri)
\le[\begin{array}{c}
\mathcal F(\xi;s)
\\
\mathcal G(\xi;s)
\end{array}\ri] &=& 0. \nonumber
\eea

Following \cite{RumanovqPII} we  impose that $\beta\in 2\N$ is an even integer and that the ratio of $B_+$ and $L_+$ has the following form
\be\label{b+}
	\frac{B_+}{L_+} := b_+(\xi;s) = -\frac{2}{\beta}\sum_{k = 1}^{\frac{\beta}2} \frac{1}{\xi - Q_{k}(s)}. 
\ee
The result of Rumanov can be summarized as follows:
\begin{theorem}[\cite{RumanovqPII}]
	If the poles $\left\{Q_{k}(s), k = 1,\ldots,\beta/2\right\}$  of $b_+$ satisfy the equation
	\be
\label{Rumanov1}
\frac{\beta^2}4 \ddot Q_k = - 2Q_k(s-Q_k^2) + \left(\frac{\beta}2-2\right) - \sum_{j \neq k}^{\beta/2} \frac {8}{(Q_k-Q_j)^3}
\ee
then there exists a Lax pair as in \eqref{RumanovLaxPair} such that $\mathcal F$ is a solution of \eqref{qPII}.
\end{theorem}
As an interesting parallel, one might say that Rumanov's result is a sort of analogue of the famous theorem by Krichever \cite{KricheverCM} stating that the poles of a rational function of the KP equation evolve according to the ``classical'' (i.e. without potential) Calogero system.

Up to a rescaling of the variables, the equation \eqref{Rumanov1} is nothing but the Hamiltonian dynamics induced by $\tilde H_{II}$ in  Takasaki's list \eqref{takalista}
Hence, from the point of view of applications to $\beta$--models, the results presented in this paper (and in particular in Section \ref{PIIsection}) provide an essential step to make  Rumanov's Lax pair effective. Indeed, the Riemann--Hilbert representation of the solutions of \eqref{Rumanov1} is needed in order to apply the Deift-Zhou non--linear steepest descent method to \eqref{RumanovLaxPair} and try to prove, for instance, the conjectured tail asymptotics of the $\beta$--Tracy--Widom distribution (see \cite{BEMN}).


From a more theoretical point of view, the study of the monodromy manifold of the second Painlev\'e\ system gives the first concrete realization of the ``quantization'' of the Stokes manifolds, in a sense demystifying the concept since it becomes simply a (structured) manifold of generalized monodromy manifold in the usual sense. The real leap to the quantum setting is provided by passing to operator-valued Stokes' parameters (see Remark \ref{remquantum}), which should be pursued on a more analytical footing.
On the other hand the quantization (this time in the sense of Reshetikhin, \cite{Reshetikhin}) of simply laced isomonodromic systems had been recently described, in full generality, in \cite{Rembado}. We plan to study in subsequent works the precise form of the monodromy manifold for the other Calogero--Painlev\'e systems, as well as their deformation quantization.\\

\section{The general scheme}\label{sectiongeneralscheme}

As explained in the introduction, our aim is to give an isomonodromic formulation of the Calogero--Painlev\'e Hamiltonian systems found by Takasaki \cite{TakasakiPaCa}. This result is achieved by performing a reduction {\it \`a la} Kazdan-Konstant-Sternberg on some matrix-valued Lax pairs for the corresponding Painlev\'e equations. For most of the cases, the Lax pairs we used are the ones found by Kawakami \footnote{Strictly speaking, for the ramified cases (PIII\textsubscript{D6}, PIII\textsubscript{D7}, PIII\textsubscript{D8}, PI), the Lax pairs of Kawakami are written for blocks of size $2\times2$, but it is easy to realise that Kawakami's formulas hold more generally for matrices of arbitrary size, and even for general non-commutative algebras.} \cite{Kawa1,Kawa2}, with the exception of the second Painlev\'e equation where we used the one found in \cite{BertolaCafasso2}, giving an isomonodromic formulation of the ``fully non--commutative'' second Painlev\'e equation introduced in \cite{RetakhRubtsov}.\\ 
For all the cases the starting point is a  Lax system of type
\be
\le\{
\begin{array}{l}
\ds 
\frac {\partial}{\partial z} \Phi (z;t) = A(z;\q , \q^{-1} , \p, t) \Phi(z;t),
\\[10pt]
\ds 
\frac {\partial}{\partial t} \Phi (z;t) = B(z;\q, \q^{-1} ,\p,t) \Phi(z;t),
\end{array}\ri.
\label{LaxPair}
\ee
where the pair of matrices $A, B$ are $2\times2$ block matrices with blocks of arbitrary size $n$. $A$ and $B$  depend rationally on the {\it spectral parameter} $z\in \mathbb{C} \mathbb P^1$ and their entries are polynomials in the $n\times n$ matrices $\{ \q ,\q^{-1}, \p \}$. The dependence on $t$ is both implicit and explicit. If a joint solution of \eqref{LaxPair} exists, then necessarily the matrices $A,B$ satisfy the {\it zero curvature equations}
\be
\label{ZC}
0\equiv [\pa_t -B, \pa_z - A] \ \ \ \Longleftrightarrow\ \ \ 
\pa_t A - \pa _z B + [A,B]\equiv 0.
\ee 
These conditions are satisfied  if $A$ and $B$ have the special form listed in Section \ref{sectionlist}; for the time being we  point out the common features to all the cases here below.  
\begin{enumerate}
\item The isomonodromic equations for the matrices $\p$ and  $\q$ can be expressed in a Hamiltonian form with  Hamiltonian function of the form $\Tr H (\p,\q, \q^{-1} )$, where $H$ is a non-commutative polynomial depending on the indicated variables. For all the cases the Hamiltonian  $H$ is a polynomial depending just on $\p$ and $\q$  except for the case PIII\textsubscript{D8}, where it is linear in $\q^{-1}$.  These provide   non--commutative versions of the Okamoto polynomial Hamiltonians. The matrices $\p,\q$ are canonically conjugate with respect to the 
symplectic structure 
\be
\omega(\p, \q) := \Tr(\d \p \wedge \d\q) = \d \theta,\qquad
\theta := \Tr \p\d \q = \sum_{j,k} p_{k,j} \d q_{j,k}. 
\ee

\item  The resulting equations for $\p, \q $ have the form 
\be
\label{QPeqs}
\dot \q  = \mathcal A(\q ,\p,t),\qquad \dot \p = \mathcal B(\q ,\p,t)
\ee
with $\mathcal A, \mathcal B$ polynomials in $\q $ (and $\q ^{-1}$ for PIII\textsubscript{D8}) and 
\bea
\hbox{$\mathcal A$ is of degree at most $1$ in $\p$ and
 $\mathcal B$ is of degree at most $2$ in $\p$.} \label{ABdegree}
\eea
\item  The ``angular momenta''
\be
\mathcal M (\p,\q):= [\p,\q]= \p\q-\q\p
\ee
are conserved quantities for the equations  \eqref{QPeqs} and the level-sets of $\mathcal M$ are the coadjoint orbits of the group $GL_n(\mathbb{C})$. This conservation law is a consequence  of the invariance of the Hamiltonian under simultaneous conjugations $$\q \mapsto C\q C^{-1}, \ \ \p \mapsto C \p C^{-1},$$ and the fact that the adjoint action of $GL_n(\mathbb{C})$ is symplectic.  
\item Reinforcing the previous point, in all cases the commutator $[\p,\q]$ has an additional interpretation in terms of the (generalized) monodromy data associated to the $z$--ODE  in \eqref{LaxPair}, which are conserved  by the isomonodromic deformation (i.e. the $t$--equation in \eqref{LaxPair}).
\end{enumerate}
\begin{remark}
	In all the cases illustrated below, the Lax pairs can be considered in the more general setting in which $\p,\q$ and $t$ are elements of a general non--commutative unital algebra, equipped with a derivation $\partial_t$ such that $\partial_t t = \1$ (see \cite{RetakhRubtsov}), and with $t$ in the center of the algebra. In particular, the features 2,3 and 4   are still valid, even if 2 has to be verified by direct computations and it does not have an interpretation in terms of symplectic actions. This more general setting will be very useful in the Section \ref{PIIsection}, where the Stokes manifold associated to the (non--commutative) second Painlev\'e equation is studied in relation with the quantization proposed in \cite{MazRou,Mazzocco13}, see Remark \ref{remquantum}.
\end{remark}
The independence of $[\p ,\q]$ on $t$ allows us to write {\it reduced} equations for the eigenvalues of $\q $ on special coadjoint orbits $\mathcal M(\p,\q)$; this reduction follows an idea that first appeared in \cite{KKS}.

\begin{lemma}\label{KKS}\cite{KKS}
	Let $\p,\q $ be  matrices  satisfying 
\be
[\p,\q ] = ig(\1 - v^T v)\ ,\qquad \mathrm{with} \quad v := (1,\ldots, 1)
\label{Orbit}
\ee
and assume that $\q $ is diagonalizable.
Then there exists an invertible matrix $C$ such that:
\begin{itemize}

   	\item C diagonalises $\q$:
   		$$\q  = C\, X\,C^{-1}, \quad X = {\rm diag}(x_1, \dots, x_n).$$
	\item The matrices $X$ and $Y := C^{-1} \p C$ satisfy the same commutation relation as $\q$ and $\p$ :		\be
			\le[Y, X  \ri] =ig(\1 - v^T v).
		\ee
\end{itemize} 
In particular, for every $j \neq k$, $j,k \in \{1,\ldots,n\}$, 
\be
\label{Yentries}
Y_{j,k} = \frac {i g}{ x_j - x_k}. 
\ee
\end{lemma}
\begin{remark}
	The diagonal entries of $Y$ are not determined by the equation \eqref{Orbit}, in the sequel they will be denoted $y_1, \ldots, y_n$.
\end{remark}
The proof is contained in \cite{KKS} and we report it here for convenience only.
\\
{\bf Proof.}
Let $\hat C$ be any diagonalising matrix for $\q $. From \eqref{Orbit} we deduce (conjugating both sides) that
\be
\le[\hat C^{-1}\p  \hat C, X  \ri] =ig\le(\1 -\hat C^{-1} v^T v\hat C \ri) = ig\le(\1 -a^T b \ri),
\ee
where $b = v\hat C$ and $a = v \hat C^{-T}$. In the (matrix) equation above, the diagonal elements on the left hand side are zero, and then we deduce that $a_j = b_j^{-1}$. Let $A := {\rm diag}(a_1,\dots, a_n)$. Then 
\be
\le\{
\begin{array}{cc}
a = v A = v \hat C^{-T}\\
  b = v A^{-1} = v\hat C
  \end{array}
  \ri. \ \ \ \Longleftrightarrow  \ \ \ \ 
  \le\{\begin{array}{cc}
  v = v (\hat{C}A)^{-T}
  \\
  v = v \hat{C}A .
  \end{array}
  \ri.
\ee
Since $\hat C$ is defined up to right multiplication by any invertible diagonal matrix $A$, we have that $ C := \hat CA$ also diagonalises $\q $ and  satisfies the  conditions of the Lemma.
\QED

The reduction to the eigenvalues now proceeds as follows; let $C=C(t)$, $\det C\equiv 1$,  be the diagonalizing matrix of Lemma \ref{KKS} and  introduce the new wave function 
$$\Psi(z;t) := \left[\begin{array}{cc} C & 0 \\ 0 & C \end{array}\right] \Phi(z;t).$$
The following proposition is an immediate consequence of the standard formulas for gauge transformations together with the definition of the matrices $X$ and $Y$.
\begin{proposition}
	The wave function $\Psi(z;t)$ is an eigenfunction of the Lax system
	\be
\le\{
\begin{array}{l}
\ds 
\frac {\pa}{\pa z} \Psi (z;t) = A(z;X , X^{-1}, Y, t) \Psi(z;t)
\\[10pt]
\ds 
\frac {\pa}{\pa t} \Psi (z;t) = \Big(B(z;X , X^{-1} ,Y ,t) - F(X,X^{-1},Y,t) \Big) \Psi(z;t)
\end{array}\ri.
\label{LaxPairII}
\ee
where $F := (C^{-1} \dot C) \otimes \1_2$. Consequently, the isomonodromic equations \eqref{QPeqs} become 
\bea
\label{Xeqs}
\dot X &=& \mathcal A(X,Y,t) + [X,F] ,\\
\label{Yeqs}
 \dot Y &=& \mathcal B(X,Y,t) + [Y,F].
\eea

\end{proposition}

It is also possible to express the entries of $F$ just in terms of the eigenvalues $\{x_1,\ldots,x_n\}$ of $X$, as explained in the following lemma.
\bl
Let C be the conjugating matrix satisfying the conditions of Lemma \ref{KKS} and $F = C^{-1}\dot{C}$. Then the entries of $F$ are given by 
\bea
(x_i - x_j)^2 F_{i,j} &=& \Big([\mathcal A(X,Y,t),X]\Big)_{i,j}, \quad i\neq j, \label{F1}\\
F_{j,j} &=& - \sum_{k: k\neq j} F_{j,k} +K, \label{F2}\\
K&:=&\frac 1  n  \sum_{\ell,m: \ell \neq m} F_{\ell,m} \label{F3}
\eea
and they are rational functions of $(x_1,\dots, x_n)$ only.
\el
{\bf Proof :}
We start from \eqref{Xeqs}. Taking the commutator $[\dot X,X]$, we get the equation
\be
\Big[X,[X,F]\Big] = [\mathcal A(X,Y),X] 
\ee
from which we deduce, since $X$ is diagonal, the formula \eqref{F1}.\\
Since $\mathcal A$ is a polynomial of first degree in $Y$ (see  \eqref{ABdegree}),  and $X$ is diagonal, the commutator in \eqref{F1} does not contain the diagonal entries of $Y$. On the other hand, the off--diagonal entries of $Y$ are expressed by \eqref{Yentries} and therefore $F_{j,k}, j\neq k$ are  rational functions of $(x_1,\dots, x_n)$ only.

In order to find the diagonal terms of $F$, we take the derivative of the commutator 
$$\frac \d{\d t}[X,Y]= [\dot X,Y] + [X,\dot Y] = 0.$$ 
Using \eqref{Xeqs} and \eqref{Yeqs} we then obtain
\be
0 &=& [\mathcal A(X,Y),X] + [Y,\mathcal B(X,Y)] + \Big( \big[[X,F], Y\big] + \big[X,[Y,F] \big] \Big). \nonumber
\ee
Now, we know that also $[\q ,\p]$ is constant and therefore $[\dot \q ,\p]+ [\q ,\dot \p]=0$; using \eqref{QPeqs} we then obtain 
\be
[\mathcal A(\q ,\p),\p] + [\p,\mathcal B(\q ,\p)] =0.
\label{219}
\ee
By conjugating  the equation \eqref{219} with $C$ we obtain  that also $ [\mathcal A(X,Y),X] + [Y,\mathcal B(X,Y)] =0$. 
Hence we conclude  that
\be
 0 = \big[[X,F], Y\big] + \big[X,[Y,F] \big] = -\big[ [Y,X], F \big] = [ig(v^Tv), F].
 \label{220}
\ee 
The off--diagonal entries of the  equation \eqref{220} give  the linear system of equations
 $$F_{i,i} + \sum_{j \neq i} F_{i,j} - F_{k,k} - \sum_{j \neq k}F_{j,k} = 0, \quad i,k = 1,\ldots,n;\, i \neq k,$$
 which  has  \eqref{F2}, \eqref{F3} for solution.
\QED
The details of the expressions of $F$ in terms of $X$ depend on the specific case considered, as will be seen in Section \ref{sectionlist}.
At the general level, however, it already follows that the eigenvalues of $X$ evolve according to a ``Calogero-Moser'' type system, in the sense specified by the following proposition:
\begin{proposition}
	The equations \eqref{Xeqs}, \eqref{Yeqs} are Hamiltonian with respect to the symplectic structure $\sum_{i = 1}^ndy_i \wedge dx_i$. Moreover, they yield a closed differential system of the second order for the eigenvalues of $X$, with poles along the diagonals $x_j=x_k,\ j\neq k$.   
\end{proposition}
{\bf Proof :}
	The fact that the equations \eqref{Xeqs}, \eqref{Yeqs} are Hamiltonian comes from the fact that the equations \eqref{QPeqs} for $\q$ and $\p$ are Hamiltonian and the action $\q \mapsto C\q C^{-1}, \ \ \p \mapsto C \p C^{-1}$ is symplectic.\\ 
From \eqref{Xeqs} and the fact that $\mathcal A$ is of first degree in $Y$, we realise that the diagonal entry $y_j := Y_{j,j}$ is an expression involving only $x_j, \dot x_j$ :
\be
y_j = W(x_j, \dot x_j).
\label{yj}
\ee
At this point $Y$ given by \eqref{Yentries} is completely determined by $X, \dot X$ and then the equation \eqref{Yeqs} yields a closed differential system of second order for the eigenvalues of $X$. The poles along the diagonals appear as a consequence of the formula \eqref{Yentries}. \QED 
Finally, when necessary (in all the cases but Painlev\'e I and II), following \cite{TakasakiPaCa}, in the next sections we  provide the explicit canonical change of variable $\{x_j,y_i\} \rightarrow \{q_j,p_j\}$ necessary to transform the equations \eqref{Xeqs},\eqref{Yeqs} in the dynamical equations for the Hamiltonian functions in the Takaksaki's list.

\section{The isomonodromic formulation of the Calogero-Painlev\'e systems}\label{sectionlist}

Following the scheme introduced in the previous Section, we explicitly display, for each of the Calogero--Painlev\'e systems,  an isomonodromic formulation. More specifically, for each case, we report:
\begin{itemize}
	\item The explicit expression of the Lax system \eqref{LaxPair}.
	\item The matrix Hamiltonian related to the equations \eqref{QPeqs}.
	\item The explicit expression of $F$.
	\item The Hamiltonian related to the equations \eqref{Xeqs}, \eqref{Yeqs}.
	\item When necessary, the change of coordinates necessary to go from the Hamiltonian of the previous point to the one of the Calogero-Painlev\'e system (see \cite{TakasakiPaCa}).
\end{itemize}

\subsection{Painlev\'e VI}\label{subsPVI}

We start from the Fuchsian system of spectral type $nn, nn, nn, n\,n-1\,1$ as written by Kawakami (up to a renaming of the constants) \cite{Kawa1}

\begin{equation}\label{PVI}
\left\{\begin{array}{lll}
	 \ds \frac{\partial \Phi}{\partial z} &=& \Bigg(\ds\frac{A_0}{z} + \ds\frac{A_1}{z - 1} + \ds\frac{A_t}{z-t} \Bigg)\Phi,\\ \\
	\ds \frac{\partial \Phi}{\partial t} &=& -\Bigg(\ds\frac{A_t}{z-t} + B\Bigg)\Phi,
\end{array}\right.
\end{equation}
where the matrices are explicitly given by
$$
A_0 := \left[ \begin {array}{cc} -1-\theta_{{t}}&\ds\frac{\q }{t} -1
\\ \noalign{\medskip}0&0\end {array} \right], 
\; 
A_1 :=  \left[ \begin{array}{cc} -\q \p + \ds\frac{1}2(k + \theta) & 1 \\
				\\
				  (\theta - \q \p)\q \p + \ds\frac{1}4(k^2 - \theta^2) & \q \p + \ds\frac{1}2(k - \theta) 
	\end{array} \right],
$$
\vspace{0.2cm}
$$
	A_t := \left[ \begin{array}{cc}  \q \p -\theta_0 & -\ds\frac{\q }t \\
					\\
				t(-\theta_0 + \p\q )\p	 & -\p\q   \end{array} \right], \;
	B := \left[ \begin{array}{cc} \ds\frac{t\Big([\q ,\p]_+ -\theta_0 \Big) + \theta \q  - [\q \p,\q ]_+}{t(t-1)}   & 0 \\
					\\
					-\theta_0 \p + \p\q \p  & 0  \end{array} \right].
$$ 
Here and below, the entries of the matrices are $(n \times n)$ blocks. When we write scalars (i.e. $0,k,t,\ldots$) we mean that this scalar is to be multiplied by the identity matrix.
$\theta_0,\theta_1,\theta_t$ and $k$ are free parameters while $\theta= \theta_0 + \theta_t + \theta_1$. 
The resulting Hamiltonian equations for $\p$ and $\q $ are given by
 	\be\label{dynamicsPVI}
	\left\{ \begin{array}{lll}
		\dot{\q } &=& \mathcal A(\q ,\p) \\
		\\
		\dot{\p} &=& \mathcal B(\q ,\p),
	\end{array}\right.
	\ee 
with
	\bea\label{defABPVI}
	t(t-1)\mathcal A(\q ,\p) &&\hspace{-15pt}:= -\theta_0 t + (\theta_0 + \theta_t)\q  + (\theta_0 + \theta_1)t\q  -\theta \q ^2 -2 \q \p\q  + t[\p,\q ]_+ - [t\p,\q ^2]_+ + [\q \p\q ,\q ]_+ \nonumber \\
	&&\\
	t(t-1)\mathcal B(\q ,\p)  &&\hspace{-15pt}:=  \ds\frac{1}4(k^2 - \theta^2) -(\theta_0 + \theta_t)\p - (\theta_0 + \theta_1)t\p + \theta[\q ,\p]_+ - t\p^2 + \nonumber\\
	 && +  t[\q ,\p^2]_+ + \p(2\q  - \q ^2)\p - [\q ,\p\q \p]_+ . \nonumber
\eea
The matrix Hamiltonian is given by
{
\be \begin{array}{lll}
t(t-1)H =\\
\Tr \Big(  \q\p\q\p\q - t \p\q^2 \p + t \p\q\p - \p\q\p\q - \theta\q\p\q + t(\theta_0 + \theta_1) \p\q + (\theta_0 + \theta_t)\p\q - \theta_0 t \p - \frac 1 4 (k^2- \theta^2) \q\Big).
\end{array}
\ee
}
The matrix $F$ reads
\be\label{eqsFPVI}
	F:= \mathrm{diag}(f_1,\dots, f_ n ) + \frac{ig}{t(t-1)}\Bigg(\frac{(t-x_i)x_i(x_i-1)+ (t-x_j)x_j(x_j-1)}{(x_i - x_j)^2} + x_i +x_j - 1  \Bigg)_{i \neq j =1}^ n ,
\ee
leading to the multi--particle Hamiltonian
\be\label{hamiltonPVI}
\begin{array}{ll}
	t(t-1)H_{VI} :=\\
	\ds\sum_{i = 1}^ n \Big[ x_i(x_i - 1)(x_i - t)y_i^2 - \Big(\theta_t x_i(x_i -1)+\theta_1x_i(x_i-t) + \theta_0(x_i-1)(x_i - t) \Big)y_i +\frac{1}4\Big(\theta^2 - {k}^2 \Big)x_i \Bigg] + \\
	 +g^2 \ds\sum_{j < k }\Bigg( \frac{x_j(x_j-1)(x_j-t) + x_k(x_k-1)(x_k-t)}{(x_j - x_k)^2} - x_j - x_k \Bigg).
\end{array}
\ee
We now recall, following Takasaki,  the change of variables  that brings the Hamiltonian \eqref{hamiltonPVI} in  ``physical'' form. We start by rewriting \eqref{hamiltonPVI} as
\bea\label{hamiltonPVI2}
	&\&t(t-1)H_{VI}  :=\nonumber \\
&\&	\ds\sum_{i = 1}^ n \Big[ x_i(x_i - 1)(x_i - t)y_i^2 - \Big(\theta_t x_i(x_i -1)+\theta_1x_i(x_i-t) + \theta_0(x_i-1)(x_i - t) \Big)y_i +\frac{1}4\Big(\theta^2 - {k}^2 + 60g^2(n-1) \Big)x_i \Bigg] + \nonumber\\
&\&	 +\ds\frac{(4g)^2}2 \ds\sum_{j < k }\Bigg( \frac{x_j(x_j-1)(x_j-t) + x_k(x_k-1)(x_k-t)}{8(x_j - x_k)^2} - 2( x_j + x_k) \Bigg).
\eea

Now introduce the Weierstrass $\wp$ function with periods $1$ and $\tau$
\be
	\wp(u) := \frac{1}{u^2} + \sum_{(m,n) \in \Z^2\backslash\{(0,0)\}} \Bigg( \frac{1}{(u + m +n\tau)^2} - \frac{1}{(m + n\tau)^2} \Bigg),
\ee
 and define
\be\label{halfperiods}
	\omega_0 := 0, \; \omega_1 := \frac{1}2, \; \omega_2 := -\frac{1 + \tau}2, \; \omega_3 := \frac{\tau}2, \quad e_i := \wp(\omega_i), \; i={1},\ldots,3.
\ee 
We perform the following change of variable\footnote{The second equation in \eqref{ell1} should be thought as an implicit relation between $t$ and $\tau$.}: 
\bea
	x_j &=& \frac{\wp(q_j) - e_1}{e_2 - e_1}, \;\qquad t = \frac{e_3 - e_1}{e_2 - e_1} \label{ell1}\\
	y_j &=& \frac{e_2 - e_1}{\wp'(q_j)}p_j + \frac{2\pi i (e_2 - e_1)^2}{\wp'(q_j)^2}f_\tau(q_j) + \frac{e_2 - e_1}2 \Big(\frac{\theta_0}{\wp(q_j) - e_1} + \frac{\theta_1}{\wp(q_j) - e_2} + \frac{\theta_t}{\wp(q_j) - e_3}\Big) \label{ell2},
\eea
where $f(u) := \frac{\wp(u)- e_1}{e_2 - e_1}$, and $f_\tau$ is its derivative with respect to $\tau$.
The resulting elliptic Hamiltonian is given by
\be
	2\pi i \tilde H = \sum_{j = 1}^n \Bigg(\frac{p_j^2}2 - \sum_{\ell = 0}^3 g_\ell \wp(q_j + \omega_\ell) \Bigg) + (4g)^2 \sum_{j < k} \Big(\wp(q_j - q_k) + \wp(q_j + q_k)\Big),
\ee
where
\be
	g_0 = \frac{1}2 ({k}^2 - 60g^2(n-1)), \; g_1 = \frac{\theta_0^2}2, \; g_2 = \frac{\theta_1^2}2, \; g_3 = \frac{{\theta_t^2+1}}2.
\ee

\subsection{Painlev\'e V}
We start from the isomonodromic system of type $(n)(n-1\, 1), nn, nn$, see equation (5.12) in \cite{Kawa1}:

\begin{equation}\label{PV}
\left\{\begin{array}{lll}
	 \ds \frac{\partial \Phi}{\partial z} &=& \Bigg(-tE_{22}+ \ds\frac{A_0}{z} + \ds\frac{A_1}{z - 1}\Bigg)\Phi,\\ \\
	\ds \frac{\partial \Phi}{\partial t} &=& B\Phi,
\end{array}\right.
\end{equation}
where the matrices are explicitly given by
$$
E_{22} :=  \left[ \begin{array}{cc} 0 & 0 \\
				\\
				  0 & 1
	\end{array} \right], \;
A_0 := \left[ \begin {array}{cc} Z_1 - Z_1 \q & Z_1 Z_2
\\ \noalign{\medskip} t - t\q  & t Z_2\end {array} \right], 
\; 
A_1 :=  \left[ \begin{array}{cc} S_1 & S_1S_2 \\
				\\
				  t\q  & t\q S_2
	\end{array} \right],
$$
\vspace{0.2cm}
$$
	B := \left[ \begin{array}{cc} 0   & \ds\frac{1}t(Z_1Z_2 + S_1S_2) \\
					\\
					1  & -z +\ds\frac{t\q  + [\p,\q ] + 1 - \theta_0 - 2\theta_1 - \theta_2}{t}  \end{array} \right].
$$
Here 
\be
\begin{array}{ll}
	Z_1 := \q \p + \theta_0 + \theta_1, & Z_2 := \ds\frac{\q ^2\p - \q \p + (\theta_0 + \theta_1)\q  - \theta_1}{t}, \\
	\nonumber\\
	S_1 := \q \p\q  - \p\q  + (\theta_0 + \theta_1)\q  + \theta_2, & S_2 := -\ds\frac{\q \p - \p + \theta_0 + \theta_1}{t}.
\end{array}
\ee
The resulting Hamiltonian equations for $\p$ and $\q $ are given by

 \be\label{dynamicsPV}
 	\left\{ \begin{array}{lll}
		\dot{\q } &=& \ds\frac{[\p,\q ^2]_+ - [\p,\q ]_+ +  t(\q ^2 - \q ) + (\theta_0 - \theta_2)\q  + \theta_2}{t}\\
		\\
		\dot{\p} &=&  \ds\frac{- [\p^2,\q ]_+ + \p^2 -t([\p,\q ]_+ + \theta_0+\theta_1) + (\theta_2 - \theta_0 + t)\p}{t},
		\end{array}\right. \\
		tH = \mathrm{Tr}\Big(\p(\p+t)\q(\q-1) + (\theta_0 - \theta_2)\p\q + \theta_2\p + (\theta_0 + \theta_1)t\q \Big).
	\ee
The matrix $F$ reads
\be\label{eqsFPV}
	F:= \mathrm{diag}(f_1,\dots, f_ n ) - \frac{ig}{t}\Bigg(\frac{x_i^2 + x_j^2 - x_i - x_j }{(x_i - x_j)^2}  \Bigg)_{i \neq j =1}^ n ,
\ee
leading to the multi--particle Hamiltonian
\be\label{hamiltonPV}
\begin{array}{ll}
	H_{V} :=
	\ds\sum_{i = 1}^ n \Bigg(\frac{x_i^2 - x_i}t \Bigg)y_i^2 + \Bigg(x_i^2 + \frac{(\theta_0 - \theta_2 - t)x_i + \theta_2}t  \Bigg)y_i+ (\theta_0 + \theta_1)x_i + \frac{g^2}t \sum_{j < k} \frac{2x_jx_k - x_j - x_k}{(x_j - x_k)^2}.
\end{array}
\ee
In order to write the physical Hamiltonian, we proceed in two steps. First we perform the change of coordinates
\be\label{change1PV}
	x_i = \frac{Q_i}{Q_i - 1}, \quad p_i = -(Q_i - 1)^2P_i - (\theta_0 + \theta_1)(Q_i - 1) 
\ee
leading to the (more familiar) Hamiltonian
\be
\begin{array}{ll}
	t \hat{H}_{V} = 
	\ds\sum_{i = 1}^ n Q_i(Q_i - 1)^2 P_i^2 + \Big((\theta_0 + 2\theta_1)Q_i^2 - (t + \theta_0 + 2 \theta_1 - \theta_2)Q_i - \theta_2  \Big)P_i + \theta_1(\theta_1 + \theta_0)Q_i \\
	+ g^2 \ds\sum_{j < k} \ds\frac{(Q_j - 1)(Q_k - 1)(Q_j + Q_k)}{(Q_j - Q_k)^2}.
\end{array}
\ee
This is the same (up to renaming the variables) as the one written in \cite{TakasakiPaCa}.
Takasaki's change of coordinates (adding a change in the time variable) reads\footnote{Here and below, actually, we write a slightly modified version of Takasaki's change of coordinates, because we prefer to write transformations that are genuinely canonical, and not canonical up to a constant, as in the Takasaki's paper (see sub-section IV.2 of \cite{TakasakiPaCa}). That is the reason why our Hamiltonians have a different normalisation with respect to the ones written in the Introduction.}
\be
	\sqrt{Q_j} = -\coth(q_j/2), \quad P_j = \frac{p_j}{\sqrt{Q_j}(Q_j - 1)} + \frac{1}2 \Bigg(\frac{\theta_2}{Q_j} - \frac{\theta_0 + 2\theta_1 + \theta_2}{Q_j - 1} + \frac{t}{(Q_j - 1)^2} \Bigg),\quad t  = {\rm e}^T,
\ee
leading to the Hamiltonian
\be
\begin{array}{ll}
	\tilde H_{V} =\\
	 \ds \sum_{i = 1}^n \Bigg( p_i^2 - \frac{1}{{32}} {\rm e}^{2T}\cosh(2q_i)  + \frac{  {\rm e}^T (\theta_0 + 2\theta_1 + \theta_2+1 )  }4 \cosh(q_i)  - \frac{\theta_0^2}4 \sinh^{-2}(q_i/2) + \frac{\theta_2^2}4 \cosh^{-2}(q_i/2) \Bigg)  \\
	 + \ds \frac{g^2}2 \sum_{j < k} \Bigg( \frac{1}{\sinh \big((q_j - q_k)/2 \big)} + \frac{1}{\sinh \big((q_j + q_k)/2 \big)} \Bigg).
\end{array}
\ee
\subsection{Painlev\'e IV}

The starting point is the isomonodromic system of spectral type $((n))((n-1\,1)),nn$ as written in \cite{Kawa1}, equation (5.16):

\begin{equation}\label{PIV}
\left\{\begin{array}{lll}
	 \ds \frac{\partial \Phi}{\partial z} &=& \left[ \begin{array}{cc} 
	 									\ds-\frac{\p\q}{z} & \q\p + \theta_0 + \theta_1 - \ds\frac{\p\q\p + \theta_0\p}{z} \\
										\\
										1 + \ds\frac{\q}z & -z + t + \ds\frac{\q\p + \theta_0}z
	 							\end{array}   \right] \Phi ,\\ \\
	\ds \frac{\partial \Phi}{\partial t} &=& \left[ \begin{array}{cc} 
	 									0 & -\q\p - \theta_0 - \theta_1\\
										\\
										-1 & z - \q - t
	 							\end{array}   \right] \Phi ,
\end{array}\right.
\end{equation}
and the resulting Hamiltonian equations for $\p,\q$ are given by

\be\label{dynamicsPIV}
\left\{\begin{array}{lll}
		\dot{\q } &=& [\p,\q]_+ -\q^2 - t\q + \theta_0 \\
		\\
		\dot{\p} &=&  [\p,\q]_+ -\p^2 + t\p + \theta_0 + \theta_1,
	\end{array}\right. \quad 
	H = \Tr \Big(\p\q(\p - \q -t) + \theta_0 \p - (\theta_0 + \theta_1)\q \Big).
\ee
The matrix $F$ reads
\be\label{eqsFPIV}
	F:= \mathrm{diag}(f_1,\dots, f_ n ) - ig\Big(\frac{x_i + x_j}{(x_i - x_j)^2}  \Big)_{i \neq j =1}^ n ,
\ee
leading to the multi--particle Hamiltonian
\be\label{hamiltonPIV}
\begin{array}{ll}
	H_{IV} :=
	\ds\sum_{i = 1}^ n x_iy_i^2 - (x_i^2 + tx_i - \theta_0)y_i - (\theta_0 + \theta_1)x_i + g^2 \sum_{j < k} \frac{x_j + x_k}{(x_j - x_k)^2}.
\end{array}
\ee
In order to write the physical Hamiltonian we perform the following change of variable
\be\label{changePIV}
	 \quad p_i := \sqrt{x_i}y_i -\frac{\sqrt{x_i}}{2}\Big(x_i + t - \frac{\theta_0}{x_i}\Big), \quad q_i := 2\sqrt {x_i}, 
\ee
bringing $H_{IV}$ into
\bea
	\tilde H_{IV} = \sum_{i = 1}^n \Bigg( p_i^2 -\frac{1}{256} q_i^6 -\frac{1}{32} t q_i^4 -\frac{1}{4} \Big(\theta_1 + \frac{\theta_0}2 + \frac{t^2}4  -\frac  12   \Big)q_i^2 -\frac{\theta_0^2}{q_i^{2}} \Bigg) + 8 g^2 \sum_{j < k} \Bigg( \frac{1}{(q_j - q_k)^2} + \frac{1}{(q_j + q_k)^2} \Bigg).\nonumber\\
\eea

\subsection{Painlev\'e III\textsubscript{D6}}

The Lax system we start from is a straightforward generalization\footnote{Here and below, when we say that the Lax system is a straightforward generalization of the one presented in \cite{Kawa2}, we mean that we use it as a system whose blocks are of arbitrary size, while the system was originally written for $2\times 2$ blocks.} of the one of spectral type $(11)_2,22,22$ written in \cite{Kawa2}, equation (3.28): 

\begin{equation}\label{PIII6}
\left\{\begin{array}{lll}
	 \ds \frac{\partial \Phi}{\partial z} &=& \left[ \begin{array}{cc} 
	 									\ds\frac{\q\p + \theta_1}{z - 1} & t - \ds\frac{\q\p\q + \theta_1\q}{z - 1}\\
										\\
										-\ds\frac{\p - 1}z + \frac{\p}{z - 1} & \ds\frac{\theta_0}z - \frac{\p\q}{z - 1}
	 							\end{array}   \right] \Phi ,\\ \\
	\ds \frac{\partial \Phi}{\partial t} &=& \left[ \begin{array}{cc} 
	 									\ds\frac{\p\q - \theta_0}{t} & z\\
										\\
										\ds\frac{1}t & -\ds\frac{\q\p + \theta_1}t
	 							\end{array}   \right] \Phi ,
\end{array}\right.
\end{equation}
and the resulting Hamiltonian equations for $\p,\q$ are given by

\be\label{dynamicsPIIID6}
\left\{ \begin{array}{lll}
		\dot{\q } &=& \ds\frac{[\p,\q^2]_+ - \q^2 + (\theta_1 - \theta_0)\q +t}{t}\\
		\\
		\dot{\p} &=&  \ds\frac{-[\p^2,\q]_+ + [\p,\q]_+ - (\theta_1 - \theta_0)\p + \theta_1}{t}
	\end{array}\right. \!\! tH = \Tr\Big( \p^2 \q^2 - (\q^2 + (\theta_0 - \theta_1)\q - t)\p - \theta_1\q \Big).
\ee
The matrix $F$ reads
\be\label{eqsFPIIID6}
	F:= \mathrm{diag}(f_1,\dots, f_ n ) - \frac{ig}{t}\Bigg(\frac{x_i^2 + x_j^2}{(x_i - x_j)^2}  \Bigg)_{i \neq j =1}^ n ,
\ee
leading to the multi--particle Hamiltonian
\be\label{hamiltonPIIID6}
\begin{array}{ll}
	H_{III_{D6}} :=
	\ds\sum_{i = 1}^ n \frac{x_i^2 y_i^2 + \big(- x_i^2 + (\theta_1 - \theta_0)x_i  +t\big)y_i -\theta_1x_i}t + \frac{2g^2}t \sum_{j < k} \frac{x_jx_k}{(x_j - x_k)^2}.
\end{array}
\ee
In order to write the physical Hamiltonian we perform the following change of variable
\be\label{changePIIID6}
	T := \ln t, \quad p_i := x_iy_i  -\frac{x_i}2  +\frac{t}{2x_i} + \frac{1}2( \theta_1 - \theta_0 ), \quad q_i := \ln x_i,
\ee
bringing $H_{III_{D6}}$ into
\bea
	\tilde H_{III_{D6}} = \sum_{i = 1}^n \Bigg( p_i^2 - \frac{1}4\big({\rm e}^{q_i} - {\rm e}^{T-q_i} \big)^2 - \frac{1}2(\theta_1 + \theta_0){\rm e}^{q_i} +\frac{( \theta_0 - \theta_1 -1 ){\rm e}^{T}}2 {\rm e}^{ -q_i} \Bigg) +  \frac{g^2}2 \sum_{j < k} \frac{1}{\sinh^2\big((q_j - q_k)/2\big)}. \nonumber \\
	{}
\eea
\subsection{Painlev\'e III\textsubscript{D7}}

The Lax system below is a straightforward generalization of the one of spectral type $(11)_2,(2)(2)$ in \cite{Kawa2}, equation (3.49): 

\begin{equation}\label{PIIID7}
\left\{\begin{array}{lll}
	 \ds \frac{\partial \Phi}{\partial z} &=& \left[ \begin{array}{cc} 
	 									\ds\frac{\q\p}{z} & \ds\frac{\q}t + 1\\
										\\
										\ds\frac{t\p}{z^2} + \ds\frac{1}z & \ds\frac{t}{z^2} + \ds\frac{-\p\q + \theta}{z}
	 							\end{array}   \right] \Phi ,\\ \\
	\ds \frac{\partial \Phi}{\partial t} &=& \left[ \begin{array}{cc} 
	 									0 & -\ds\frac{\q}t\\
										\\
										-\ds\frac{\p}z & -\frac{1}z
	 							\end{array}   \right] \Phi,
\end{array}\right.
\end{equation}
and the resulting Hamiltonian equations for $\p,\q$ are given by

\be\label{dynamicsPIIID7}
\left\{\begin{array}{lll}
		\dot{\q } &=& \ds\frac{2\q\p\q - \theta\q + t}{t}\\
		\\
		\dot{\p} &=&  -\ds\frac{2\p\q\p - \theta\p + 1}{t},
\end{array}\right. \quad tH = \Tr( {\p\q\p\q} - \theta \p\q + t\p + \q).
	\ee
The matrix $F$ reads
\be\label{eqsFPIIID7}
	F:= \mathrm{diag}(f_1,\dots, f_ n ) - \frac{2ig}{t}\Bigg(\frac{x_ix_j}{(x_i - x_j)^2}  \Bigg)_{i \neq j =1}^ n ,
\ee
leading to the multi--particle Hamiltonian
\be\label{hamiltonPIIID7}
\begin{array}{ll}
	H_{III_{D7}} :=
	\ds\sum_{i = 1}^ n \frac{x_i^2 y_i^2 + \big(t - \theta x_i \big)y_i +x_i}t + \frac{2g^2}t \sum_{j < k} \frac{x_jx_k}{(x_j - x_k)^2}.
\end{array}
\ee
In order to write the physical Hamiltonian one has to perform the following change of variables
\be\label{changePIIID7}
	T := \ln t, \quad p_i := x_iy_i + \frac{t}{2x_i} - \frac{\theta}2, \quad q_i := \ln x_i,
\ee
bringing $H_{III_{D7}}$ into
\be
	\tilde H_{III_{D7}} = \sum_{i = 1}^n  \Bigg( p_i^2 + {\rm e}^{q_i} + \frac{(\theta {-1}) {\rm e}^{T}}2 {\rm e}^{ - q_i} - \frac{1}4 {\rm e}^{2T - 2q_i} \Bigg) + \frac{g^2}2 \sum_{j < k} \frac{1}{\sinh^2\big((q_j - q_k)/2\big)}.
\ee

\subsection{Painlev\'e III\textsubscript{D8}}

The Lax system, generalising the one of spectral type $(2)_2,(11)_2$, equation (3.57) in \cite{Kawa2}, reads 

\begin{equation}\label{PIIID8}
\left\{\begin{array}{lll}
	 \ds \frac{\partial \Phi}{\partial z} &=& \left[ \begin{array}{cc} 
	 									\ds\frac{\q\p}{z} & -\ds\frac{\q}z + 1\\
										\\
										\ds-\frac{t\q^{-1}}{z^2} + \ds\frac{1}z & - \ds\frac{\p\q + 1}{z}
	 							\end{array}   \right] \Phi ,\\ \\
	\ds \frac{\partial \Phi}{\partial t} &=& \left[ \begin{array}{cc} 
	 									0 & \ds\frac{\q}t\\
										\\
										\ds\frac{\q^{-1}}z & 0
	 							\end{array}   \right] \Phi,
\end{array}\right.
\end{equation}
and the resulting Hamiltonian equations for $\p,\q$ are given by

\be\label{dynamicsPIIID8}
\left\{ \begin{array}{lll}
		\dot{\q } &=& \ds\frac{2\q\p\q + \q}{t}\\
		\\
		\dot{\p} &=&  -\ds\frac{2\p\q\p + \p - 1}{t}-\q^{-2}
\end{array}\right. \quad tH = \Tr( {\p\q\p\q} + \p\q -\q - t\q^{-1}).
\ee
The matrix $F$ reads
\be\label{eqsFPIIID8}
	F:= \mathrm{diag}(f_1,\dots, f_ n ) - \frac{2ig}{t}\Bigg(\frac{x_ix_j}{(x_i - x_j)^2}  \Bigg)_{i \neq j =1}^ n ,
\ee
leading to the multi--particle Hamiltonian
\be\label{hamiltonPIIID8}
\begin{array}{ll}
	H_{III_{D8}} :=
	\ds\sum_{i = 1}^ n \Bigg( \frac{x_i^2 y_i^2}t + \frac{x_iy_i - x_i - tx_i^{-1}}t \Bigg) + \frac{2g^2}t \sum_{j < k} \frac{x_jx_k}{(x_j - x_k)^2}.
\end{array}
\ee

In order to write the physical Hamiltonian one has to perform the following change of variable
\be\label{changePIIID8}
	T := \ln t, \quad p_i := x_iy_i + \frac{1}2, \quad q_i := \ln x_i,
\ee
bringing $H_{III_{D8}}$ into
\be
	\tilde H_{III_{D8}} = \sum_{i = 1}^n \Bigg( p_i^2 - {\rm e}^{q_i} -  {\rm e}^{T-q_i} \Bigg) + \frac{g^2}2 \sum_{j < k} \frac{1}{\sinh^2\big((q_j - q_k)/2\big)}.
\ee
\subsection{Painlev\'e II}\label{SectionPII}

The Lax system, which is a particular case of the one used in \cite{BertolaCafasso2}, is the following:

\begin{equation}\label{PII}
\left\{\begin{array}{lll}
	 \ds \frac{\partial \Phi}{\partial z} &=& \left[ \begin{array}{cc} 
	 									i \ds\frac{z^2}2 + i\q^2 + i\frac{t}2 & z\q - i\p -  \ds\frac{\theta}z\\
										\\
										z\q + i\p - \ds\frac{\theta}z & - i \ds\frac{z^2}2 - i\q^2 - i\ds\frac{t}2
	 							\end{array}   \right] \Phi ,\\ \\
	\ds \frac{\partial \Phi}{\partial t} &=& \left[ \begin{array}{cc} 
	 									i\ds\frac{z}2 & \q\\
										\\
										\q & -i\ds\frac{z}2
	 							\end{array}   \right] \Phi ,
\end{array}\right.
\end{equation}
and the resulting Hamiltonian equations for $\p,\q$ are given by

\be\label{dynamicsPII}
\left\{\begin{array}{lll}
		\dot{\q } &=& \p\\
		\\
		\dot{\p} &=&  2\q^3 + t\q + \theta
\end{array}\right. \quad H = \Tr\Bigg(\frac{\p^2}2 - \frac{1}2 \Big(\q^2 + \frac{t}2 \Big)^2 - \theta \q \Bigg).
\ee
The matrix $F$ reads
\be\label{eqsFPII}
	F:= \mathrm{diag}(f_1,\dots, f_ n ) - \Bigg(\frac{ig}{(x_i - x_j)^2}  \Bigg)_{i \neq j =1}^ n ,
\ee
leading to the multi--particle Hamiltonian
\be\label{hamiltonPII}
\begin{array}{ll}
	H_{II} :=
	\ds\sum_{i = 1}^ n \Bigg( \frac{y_i^2}2 -\frac{1}2\Big(x_i^2 + \frac{t}2\Big)^2 - \theta x_i \Bigg) + \sum_{j < k} \frac{g^2}{(x_j - x_k)^2}.
\end{array}
\ee
This is already in the Takasaki's form, and no additional canonical transformation is required.
\subsection{Painlev\'e I}

The Lax system reads 

\begin{equation}\label{PI}
\left\{\begin{array}{lll}
	 \ds \frac{\partial \Phi}{\partial z} &=& \left[ \begin{array}{cc} 
	 									\p & z - \q\\
										\\
										z^2 + z\q + \q^2 + \ds\frac{t}2 & - \p
	 							\end{array}   \right] \Phi ,\\ \\
	\ds \frac{\partial \Phi}{\partial t} &=& \left[ \begin{array}{cc} 
	 									0 & \ds\frac{1}2\\
										\\
										\ds\frac{z}2 + \q & 0
	 							\end{array}   \right] \Phi ,
\end{array}\right.
\end{equation}
and the resulting Hamiltonian equations for $\p,\q$ are given by

\be\label{dynamicsPI}
\left\{ \begin{array}{lll}
		\dot{\q } &=& \p\\
		\\
		\dot{\p} &=&  \ds\frac{3}2 \q^2 + \ds\frac{t}4 
\end{array}\right. \quad
H = \mathrm{Tr}\Big(\frac{\p^2}2 - \frac{\q^3}2 - \frac{t\q}4 \Big).
\ee
The matrix $F$ reads
\be\label{eqsFPI}
	F:= \mathrm{diag}(f_1,\dots, f_ n ) - \Bigg(\frac{ig}{(x_i - x_j)^2}  \Bigg)_{i \neq j =1}^ n ,
\ee
leading to the multi--particle Hamiltonian
\be\label{hamiltonPI}
\begin{array}{ll}
	H_{I} :=
	\ds\sum_{i = 1}^ n \Bigg( \frac{y_i^2}2 -\frac{x_i^3}2 - \frac{t x_i}4\Bigg) + \sum_{j < k} \frac{g^2}{(x_j - x_k)^2}.
\end{array}
\ee
This is already in the Takasaki's form, and no additional canonical transformations is required.
\section{A case study: the second Calogero--Painlev\'e system}\label{PIIsection}

In this section as an example, we study the monodromy data manifold associated to the second Calogero--Painlev\'e system \eqref{PII}--\eqref{hamiltonPII}. Quite interestingly this analysis  leads to a non--commutative deformation of the usual cubic defining the monodromy surface of the PII equation. This deformation shares many similarities with the quantization procedure proposed in \cite{MazRou,Mazzocco13}. In the second part we show how, using Schlesinger transformations, we can compute solutions of the Calogero--Painlev\'e system  \eqref{PII}--\eqref{hamiltonPII} (for particular values of the interaction constant $g$) out of $n$ (different) decoupled scalar solutions of the Painlev\'e II equation. 

\subsection{Formal solutions and generalized monodromy manifold}

In this subsection we will study the monodromy associated to the system \eqref{PII} under a more general setting than the one of the previous sections. Namely, we just assume that  $\p$ and $\q$ are symbols of a general unital and non--commutative algebra $\cA$ over $\mathbb{C}$, equipped with a derivation $\partial_t$ such that $\partial_t t = 1$and $t$ is in the center of the algebra.
 We will also denote
$$A(z) :=  \left[ \begin{array}{cc} 
	 									i \ds\frac{z^2}2 + i\q^2 + i\frac{t}2 & z\q - i\p -  \ds\frac{\theta}z\\
										\\
										z\q + i\p - \ds\frac{\theta}z & - i \ds\frac{z^2}2 - i\q^2 - i\ds\frac{t}2
	 							\end{array}   \right] .$$
In the following, given an element $a \in \mathcal A$ and $G := \Big(G_{i,j} \Big)_{i,j = 1}^2$ a matrix with complex coefficients,  we indicate with the tensor product $A \otimes B$ the matrix (with entries in the algebra $\mathcal A$)
$$
	a \otimes G := \left(\begin{array}{cc}
					a G_{1,1} & aG_{1,2}\\
					aG_{2,1} & aG_{2,2}
				\end{array}\right).
$$
The symbols $\wh \s_i, \, i = 1,\ldots,3$, will denote the matrices $\1 \otimes \sigma_i $, where $\sigma_i$ are the usual Pauli matrices.\\
The first step is finding, in the spirit of \cite{JMU1, Wasow}, a formal solution $\Phi_{\form}$ (around infinity) of the equation
\be\label{z-equation}
	\frac{d}{d z}\Phi(z) = A(z)\Phi(z)
\ee
The following is a generalization, to the non--commutative case, of the Proposition 2.2 in \cite{JMU1}.
\bp
	Assume that the operator $\Id + \ad_{[\q,\p]}$ is invertible in $\cA$. Then the equation \eqref{z-equation} admits a (unique) formal solution (around infinity) of the form 
	\be
\Phi_{\form}(z) =\le(\1 + \frac {Y_1^{(3)} \ot \s_3 + Y_{1}^{(2)} \ot \s_2}z + \frac {Y_2^{(0)} \ot \1 + Y_2^{(1)} \s_1}{z^2} + \dots\ri) z^{[\q,\p]\ot \1} {\rm e}^{\frac i  2\le(\frac {z^3}3 + tz \ri) \wh \s_3}
\label{PhiForm}
\ee
where the form of general coefficient in the $z^{-1}$ expansion depends on the parity of $j$ as illustrated by the first two terms, and
\be\label{hamiltonian}
	Y_1^{(3)} + \ad_{ [ \q,\p]} Y_1^{(3)} = -i\le(\q^4 - \p^2 + t\q^2 + 2\theta \q\ri), \quad \quad Y_1^{(2)} = -\q. 
\ee
\ep
{\bf Proof.} 
We start observing that, because of the symmetry $-A(-z) = \wh \s_1 A(z) \wh \s_1$, a solution should be found in the following form
\be
\Phi_{\form}(z) =\le(\1 + \frac {Y_1^{(3)} \ot \s_3 + Y_{1}^{(2)} \ot \s_2}z + \frac {Y_2^{(0)} \ot \1 + Y_2^{(1)} \s_1}{z^2} + \dots\ri) J(z),
\label{form1}
\ee
where $J(z)$ is block--diagonal but, in general, with exponential behaviour.\\
Denoting by $Y(z)$ the formal series in the bracket of \eqref{form1} we observe that it can be uniquely factored in the form 
\bea 
	Y(z) &=& F(z) D(z), \quad {\mathrm{where}}\\
	F(z) &=& \1 + \sum_{j=1}^\infty F_j z^{-j}\\
	D(z) &=& \1 + \sum D_{j} z^{-j}
\eea
where $F_j$ are block-off-diagonal, while $D_j$ are block-diagonal. Note that $D$ is a formally invertible matrix-valued series.\\
In other words, we look for a solution of the form 
\be
\Phi_{\form}(z)  = F(z) D(z) J(z).
\ee

The equation \eqref{z-equation}  can then be re-written as 
\be
F'  + F D' D^{-1} =  A F   + F D J' J^{-1}  D^{-1}\label{master}
\ee 
(here the prime denotes the derivatives with respect to $z$, and the dependance by $z$ is not explicitly written).
Taking the block--diagonal part of \eqref{master} (denoted here with the subscripts $_{_D}$) we obtain (re-arranging the terms) the equation
\bea
DJ'J^{-1}D^{-1}  -  D'D^{-1} = (AF)_{_D}.  \label{MasterJ}
\eea
Hence we deduce that conclude that $J'J^{-1}$ must be a polynomial of degree at most $2$ plus a term proportional to $z^{-1}$; the coefficients of the powers $z^{-j}, j\geq 2$ in \eqref{MasterJ}  determine recursively the operators $D_j$. Indeed, they have the form
\be\label{DRecursion}
D_j  +  \bigg[\big[ \q, \p\big], D_j\bigg] = \hbox{ expression containing $D_{j-\ell}$, $\ell\geq 1$},
\ee 
which can be solved (under our assumption) for $D_j$.\\ 

The block-off-diagonal part (denoted with the subscript $_{_{OD}}$) gives (using \eqref{MasterJ}) 
\be
F' = (AF)_{_{OD}}  - (AF)_{_D} F\label{MasterF},
\ee
which recursively determines uniquely the operator coefficients $F_j$ (the system for the off--diagonal parts of $F_j$ is quadratic but of triangular nature). \\
Finally, the precise form of the exponential terms in $J(z)$ and the equation  \eqref{hamiltonian} are obtained by direct computation of the first few coefficients of \eqref{MasterJ}.
\QED 

We now proceed to the study of the Stokes data associated to the equation \eqref{z-equation}. 
The situation does not  differ significantly from the scalar (commutative)  case and we can use much of the logic that was used by Flaschka and Newell in \cite{FN}, in particular the authors' discussion on pages 83--85. One difference  in our case is that, due to the term $z^{[\q,\p]\ot \1}$, there is also an exponent of  formal monodromy around $z=\infty$, in addition to the ordinary monodromy around $z=0$ of an arbitrary solution of \eqref{z-equation}.\\

We thus consider the eight rays $R_j, j = 0, \ldots 7$, defined by
$$R_0= \R_+, \quad R_j= {\rm e}^{k i\pi/3 + i\pi/6}\R_+, \; k=1,2,3, \quad R_{j+4}=-R_j,,\; j = 0,\ldots,4.$$
We denote $S_\nu$ the sector between the rays $R_{\nu}$ and $R_{\nu + 1}$.\\ 
We want to construct a piecewise analytic solution $\Phi$ of \eqref{z-equation} consisting of an analytical solution $\Phi_\nu$ for each sector $S_\nu$, satisfying the symmetry 
\be
\Phi(-z) = \wh \s_1 \Phi(z) \wh \s_1,  \label{symmPhi}
\ee
and the asymptotic condition \eqref{PhiForm2} below, which is a slight modification of \eqref{PhiForm}:
\be
\Phi(z) \sim \le(\1 + \frac {Y_1^{(3)}\ot \s_3 + Y_1^{(2)}\ot \s_2}z  + \mathcal O(z^{-2}) \ri) {\rm e}^{\le(\ln z + i\pi  \epsilon\ri)[\q,\p]\ot \1} {\rm e}^{\frac i  2\le(\frac {z^3}3 + tz \ri) \wh \s_3}.
\label{PhiForm2}
\ee
Here $\epsilon = 1$ in the lower half plane an  $\epsilon =0$ in the upper, and we use the principal value $\arg (z)\in (-\pi, \pi]$. 
 
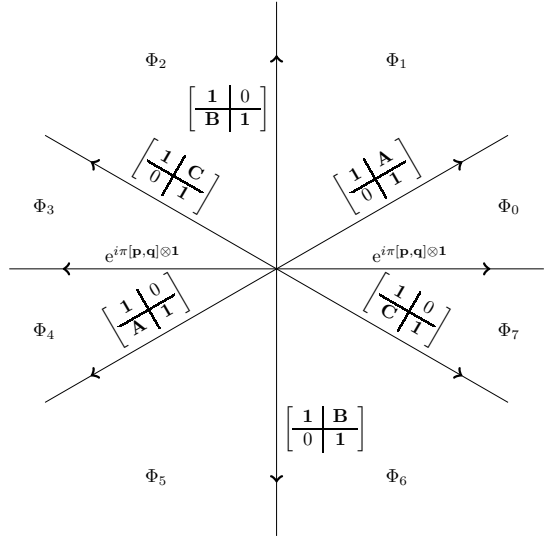
\begin{figure}
\begin{center}
 \resizebox{0.4\textwidth}{!}{
 \begin{tikzpicture}[scale=2.5]
\draw [  postaction={decorate,decoration={markings,mark=at position 0.8 with {\arrow[line width=1.5pt]{>}}}}] (0,0) to node [pos=0.5, above] {${\rm e}^{i\pi [\p,\q]\ot \1}$} (2,0);
\draw [  postaction={decorate,decoration={markings,mark=at position 0.8 with {\arrow[line width=1.5pt]{>}}}}] (0,0) to node [pos=0.5, above] {${\rm e}^{i\pi [\p,\q]\ot \1}$} (-2,0);
\draw [  postaction={decorate,decoration={markings,mark=at position 0.8 with {\arrow[line width=1.5pt]{>}}}}] (0,0) to node [pos=0.5, above, sloped] {$\le[\begin{array}{c|c}
\1 & \A\\
\hline
0&\1
\end{array}\ri]$} (30:2);
\draw [  postaction={decorate,decoration={markings,mark=at position 0.8 with {\arrow[line width=1.5pt]{>}}}}] (0,0) to node [pos=0.6, left] {$\le[\begin{array}{c|c}
\1 & 0\\
\hline
\B &\1
\end{array}\ri]$} (90:2);
\draw [  postaction={decorate,decoration={markings,mark=at position 0.8 with {\arrow[line width=1.5pt]{>}}}}] (0,0) to node [pos=0.5, above, sloped] {$\le[\begin{array}{c|c}
\1 & \C\\
\hline
0 &\1
\end{array}\ri]$} (150:2);
\draw [  postaction={decorate,decoration={markings,mark=at position 0.8 with {\arrow[line width=1.5pt]{>}}}}] (0,0) to node [pos=0.5, above, sloped] {$\le[\begin{array}{c|c}
\1 & 0\\
\hline
\A &\1
\end{array}\ri]$} (210:2);
\draw [  postaction={decorate,decoration={markings,mark=at position 0.8 with {\arrow[line width=1.5pt]{>}}}}] (0,0) to node [pos=0.6, right] {$\le[\begin{array}{c|c}
\1 & \B\\
\hline
0 &\1
\end{array}\ri]$} (270:2);
\draw [  postaction={decorate,decoration={markings,mark=at position 0.8 with {\arrow[line width=1.5pt]{>}}}}] (0,0) to node [pos=0.5, above, sloped] {$\le[\begin{array}{c|c}
\1 & 0\\
\hline
\C &\1
\end{array}\ri]$} (330:2);

\node at (15:1.8) {$\Phi_0$};
\node at (60:1.8) {$\Phi_1$};
\node at (120:1.8) {$\Phi_2$};
\node at (165:1.8) {$\Phi_3$};
\node at (195:1.8) {$\Phi_4$};
\node at (240:1.8) {$\Phi_5$};
\node at (300:1.8) {$\Phi_6$};
\node at (345:1.8) {$\Phi_7$};

\end{tikzpicture}}
\end{center}
\caption{The sectors for the sectionally analytic solution of \eqref{z-equation}.}
\label{Stokes2}
 \end{figure}

 \bt
\label{thmqstokes}
For each of the sector $S_{\nu}, \; \nu = 0,\ldots,7$, there exists a unique solution $\Phi_{\nu}$ such that the piecewise analytic solution $\Phi = \{\Phi_\nu,\, \nu = 0,\ldots,7\}$ of \eqref{z-equation} satisfies the asymptotic \eqref{PhiForm2} and the symmetry condition \eqref{symmPhi}. Moreover, $\Phi$ is the unique solution of the Riemann--Hilbert problem with the jump matrices indicated in the Figure \ref{Stokes2} where the (matrix) Stokes operator $\A,\B,\C$ satisfy the relations 
\bea
(\A+\C+\A\B\C) Q+ Q^{-1}\B &\&= 2i\sin(\pi \theta),
\label{Rel1}
\\
(\A\B+\1)Q -Q^{-1} (\B\A+\1)&\&=0,
\label{Rel2}
\\
\C Q\A-\A Q^{-1}\C +Q-Q^{-1} &\& =0, \label{redu}
\\
(\B\C+\1)Q -Q^{-1} (\C\B+\1)  &\&=0,
\label{Rel3}
\\
\B Q +Q^{-1} (\A+\C+\C \B \A) &\& = 2i\sin(\pi \theta),
\label{qstokes}
\eea
and $Q := {\rm e}^{i\pi [\p,\q]}$.
\et

Before the proof, we make three remarks. 

\br
\label{remclassic}
The relations \eqref{qstokes} should be viewed as a non-commutative version of the relation (3.24) of \cite{FN} for the scalar Stokes' parameters $a,b,c$ (here below we follow their notations):
\be
\label{FNstokes}
a+b+c + abc= -2i\sin(\pi \nu).
\ee
Indeed, if  $Q=\pm \1$ we can see easily that $\A, \B, \C$ all commute between each other and the first (or last) equation in \eqref{qstokes} exactly reduces to \eqref{FNstokes} with the opposite sign of $\theta=-\nu$ for $Q=\1$ and $\theta=\nu$  for the case $Q=-\1$. 
\er

The relations \eqref{qstokes}, as per Remark above, become commutative, of course, if $[\p,\q]=0$ but  also, and quite interestingly, in the matricial setting when the spectrum of the commutator $[\p,\q]$ consists of only even integers or only odd integers. This observation is the hinge of the construction in Section \ref{specsols}.

\br
\label{remquantum}
If $[\p,\q]= i\hbar$ is a multiple of the identity (which can only occur in a genuinely operatorial setting) then the relations \eqref{qstokes} define a ``quantum'' version of \eqref{FNstokes}. 
In this case, the computation of the formal expansion \eqref{PhiForm} of the function $\Phi$ is simplified and the operator $Y_1^{(3)}$ in \eqref{hamiltonian} is directly the quantum Hamiltonian of q(uantum)PII, upon identifying (as usual) $\p$ with $i\hbar \frac{\partial}{\partial \q}$. 
In this case, the relations \eqref{Rel2}, \eqref{redu}, \eqref{Rel3} give the {\it equitable} representation of $U_Q(\mathfrak sl_2)$ ($Q$ is now  a multiple of the identity) \cite{ItoTerwilliger}. Interestingly, the same relations had been recovered by one of the authors and M. Mazzocco in \cite{MazRou} as a result of the quantization of the monodromy manifold associated to the Flashka--Newell Lax pair for Painlev\'e II (see also \cite{Mazzocco13} for the case of the Jimbo--Miwa Lax pair).
\er

\br
\label{StokesetWilson}
The Riemann--Hilbert correspondence associate to each solution of the Hamiltonian equations related to \eqref{hamiltonPII} a point in the Stokes manifold $\mathcal S_n$ defined by the equations \eqref{Rel1}--\eqref{qstokes}:
\be
	\{x_1(t),\ldots,x_n(t)\} \longleftrightarrow \{\A,\B,\C \}.
\ee 
Hence, much in the same way as Okamoto \cite{OkamotoSurfaces} did for classical Painlev\'e equations (see also \cite{SaitovanderPut}), we can identify $\mathcal S_n$ with the space of initial conditions for the second Calogero--Painlev\'e system (with $n$ particles). For the case of the rational Calogero--Moser system, the corresponding object (the space of the initial conditions) is the so--called Kazdan-Konstant-Sternberg space $\mathcal C_n$, defined as the space of all pairs of $(n\times n)$ complex matrices $(\p,\q)$ such that 
\be
	[\p,\q] + \1_n \quad \textrm{has rank 1},
\ee
modulo the action of the linear group by simultaneous conjugations. $\mathcal C_n$ had been defined in \cite{KKS} and identified in \cite{WilsonAdelic} with the so--called adelic Grassmannian (more precisely, this latter decomposes into a disjoint union of pieces isomorphic to $\mathcal C_n, \; n \in \mathbb Z_+$). The main difference between the Painlev\'e--Calogero system and the rational Calogero model is that, in this second case, the Hamiltonian is autonomous and hence each cell $\mathcal C_n$ coincides with the phase space of the system (with $n$ particles). Moreover, essentially because of integrability, the flow is linear on these cells. In the Calogero--Painlev\'e case, since the Hamiltonian depends explicitly on the time, the identification between the space of initial conditions and the phase space is missing in the first place. Nevertheless, we believe that the geometry of the manifolds $\mathcal S_n$ plays an important role in the study of the Calogero--Painlev\'e systems and deserves a closer investigation. For instance it would be interesting, as it had been done in \cite{ShiodaTakano} and \cite{cubicpencils} for the case of Painelv\'e equations, to use the properties of $\mathcal S_n$ to recover the Hamiltonian \eqref{hamiltonPII}.
\er

\noindent {\bf Proof of Thm. \ref{thmqstokes}.}\\
\indent The proof follows closely the scalar case (pag. 84-85 in \cite{FN}), hence we will limit ourselves to explain how the arguments of Flashka and Newell should be modified to replace the relation \eqref{FNstokes} with the non--commutative analogue \eqref{Rel1},\eqref{Rel2},\eqref{redu},\eqref{Rel3} and \eqref{qstokes}.\\ 
Let $\Phi_0(z)$ be one of the solutions of \eqref{z-equation} with the asymptotic behaviour \eqref{PhiForm2} in the sector $\mathcal S_0$; its analytic continuation around a simple loop encircling the origin yields the same solution up to right multiplication by the {\it monodromy} matrix $M_0$. On the other hand, by  using the relations between consecutive sectors of the various sectional solutions $\Psi_\nu$, we obtain the {\it monodromy relation}
\bea
\le[\begin{array}{cc} \1 & 0 \\
\A & \1
\end{array}  \ri]\le[\begin{array}{cc} \1& \B\\ 0  & \1
\end{array}  \ri]\le[\begin{array}{cc} \1 & 0  \\ \C &\1
\end{array}  \ri]
\le[\begin{array}{cc}
Q&0
\\
0&Q
\end{array}\ri]\le[\begin{array}{cc} \1 & \A\\
0 & \1
\end{array}  \ri]\le[\begin{array}{cc} \1&0\\ \B & \1
\end{array}  \ri]\le[\begin{array}{cc} \1 & \C \\ 0&\1
\end{array}  \ri] \le[\begin{array}{cc}
Q&0
\\
0&Q
\end{array}\ri]= M_0^{-1} 
\label{mono1}
\eea
Now observe that, since $M_0$, must be in the same conjugacy class as ${\rm e}^{-2i\pi \theta\wh \s_1}$, we have
\be
M_0 = \le[\begin{array}{cc} {\a} &\b\\ \gamma & \delta
\end{array}  \ri]{\rm e}^{-2i\pi \theta \wh \s_3} 
\le[\begin{array}{cc}  \wt  \a &  \wt  \b\\ \wt   \gamma & \wt   \delta
\end{array}  \ri]
\ee
where $\alpha, \wt \alpha,$ etc. are matrices of size $n$ and satistying 
\be
\le[
\begin{array}{cc}
\alpha & \beta
\\
\gamma  & \delta
\end{array}\ri] 
\le[
\begin{array}{cc}
\wt  \alpha & \wt \beta
\\
\wt \gamma  & \wt \delta
\end{array}\ri]=
\le[
\begin{array}{cc}
\1 & 0 \\ 0 & \1
\end{array}
\ri].
\label{inverse}
\ee
The relation \eqref{mono1} does not use the symmetry \eqref{symmPhi} and thus it misses some information. Using \eqref{symmPhi}, we observe that \eqref{mono1} can be written as $G^2 = M_0^{-1}$ with 
\be
G := \le[\begin{array}{cc} \1 & \A\\
0 & \1
\end{array}  \ri]\le[\begin{array}{cc} \1&0\\ \B & \1
\end{array}  \ri]\le[\begin{array}{cc} \1 & \C \\ 0&\1
\end{array}  \ri]  
\le[\begin{array}{cc}
Q&0
\\
0&Q
\end{array}\ri]\wh \s_1.
\ee
Therefore, we can rewrite \eqref{mono1} in a ``square-root'' form 
\bea
\le[\begin{array}{cc} \1 & \A\\
0 & \1
\end{array}  \ri]\le[\begin{array}{cc} \1&0\\ \B & \1
\end{array}  \ri]\le[\begin{array}{cc} \1 & \C \\ 0&\1
\end{array}  \ri]  
\le[\begin{array}{cc}
Q&0
\\
0&Q
\end{array}\ri]
=
\le[\begin{array}{cc} {\a} &\b\\ \gamma & \delta
\end{array}  \ri]{\rm e}^{i\pi \theta \wh \s_3} 
\le[\begin{array}{cc}  \wt  \a &  \wt  \b\\ \wt   \gamma & \wt   \delta
\end{array}  \ri]\wh \s_1 .
\label{rel1bis}
\eea
We also need its inverse
\bea
\le[\begin{array}{cc} Q^{-1}&0\\0&Q^{-1}
\end{array}  \ri]
\le[\begin{array}{cc} \1 & -\C\\ 0&\1
\end{array}  \ri]\le[\begin{array}{cc} \1&0\\-\B & \1
\end{array}  \ri]\le[\begin{array}{cc} \1 & -\A\\
0 & \1
\end{array}  \ri]=
\wh \s_1
\le[\begin{array}{cc} {\a} &\b\\ \gamma & \delta
\end{array}  \ri]{\rm e}^{-i\pi \theta \wh \s_3} 
\le[\begin{array}{cc}  \wt  \a &  \wt  \b\\ \wt   \gamma & \wt   \delta
\end{array}  \ri].
\label{rel2bis}
\eea
Expanding both sides of \eqref{rel1bis}, \eqref{rel2bis} we obtain the (non--commutative) relations
\bea
\le[\begin{array}{cc} 
Q +  \A\B Q & \A Q + \C Q +  \A\B\C Q
\\
\B Q & Q +  \B \C Q
\end{array}  \ri] &=& \le[\begin{array}{cc} {\rm e}^{i\pi \theta} \alpha \wt  \beta  - 
{\rm e}^{-i\pi \theta} \beta \wt  \delta
& 
{\rm e}^{i\pi \theta} \alpha \wt  \alpha  - 
{\rm e}^{-i\pi \theta} \beta \wt  \gamma
\\
{\rm e}^{i\pi \theta} \gamma \wt  \beta  - 
{\rm e}^{-i\pi \theta} \delta  \wt  \delta
&
{\rm e}^{i\pi \theta} \gamma \wt  \alpha  - 
{\rm e}^{-i\pi \theta} \delta  \wt  \gamma
\end{array}  \ri]\label{nc-1}
\\
\le[\begin{array}{cc}Q^{-1} (\C \B + \1)  &  -Q^{-1}(\A + \C + \C \B\A)
\\
-Q^{-1}\B  &Q^{-1}  (\B\A+\1)
\end{array}  \ri] &=& \le[\begin{array}{cc} -{\rm e}^{i\pi \theta} \delta \wt  \gamma  
+
{\rm e}^{-i\pi \theta} \gamma \wt  \alpha
& 
{\rm e}^{-i\pi \theta} \gamma \wt  \beta
-{\rm e}^{i\pi \theta} \delta \wt  \delta 
\\
{\rm e}^{-i\pi \theta} \alpha  \wt  \alpha
-{\rm e}^{i\pi \theta} \beta \wt  \gamma 
&
{\rm e}^{-i\pi \theta} \alpha  \wt  \beta
-{\rm e}^{i\pi \theta} \beta \wt  \delta  
\end{array}  \ri].\label{nc-2}
\eea
Linear combinations of \eqref{nc-1},\eqref{nc-2} yield
\bea
(\A+\C+\A\B\C)Q + Q^{-1} \B  &=& 2i\sin(\pi \theta) \le( \alpha \wt  \alpha + \beta \wt  \gamma\ri)\mathop{=}^{\hbox{\tiny\eqref{inverse}}_{11}} 2i\sin(\pi \theta) \nonumber
\\
 (\A\B+\1) Q -Q^{-1}  (\B\A+\1) &=& 2i\sin(\pi \theta) \le(\alpha \wt  \beta + \b\wt  \delta\ri) \mathop{=}^{\hbox{\tiny\eqref{inverse}}_{12}} 0 \nonumber
\\
(\B\C+\1)Q -Q^{-1}  (\C\B+\1) &=& 2i\sin(\pi \theta) \le( \gamma \wt  \alpha + \delta \wt  \gamma\ri) \mathop{=}^{\hbox{\tiny\eqref{inverse}}_{21}}0 \nonumber
\\
\B Q + Q^{-1} (\A+\C+\C\B\A)  &=& 2i\sin(\pi \theta) \le( \gamma \wt  \beta + \delta \wt  \delta\ri)\mathop{=}^{\hbox{\tiny\eqref{inverse}}_{22}} 2i\sin(\pi \theta), \nonumber
\eea
which together with \eqref{inverse} provide  the  relations \eqref{Rel1},\eqref{Rel2},\eqref{Rel3} and \eqref{qstokes}.  The relation \eqref{redu} is obtained considering the expression $\eqref{Rel1}\A- \A \eqref{qstokes}$ and simplifying it using \eqref{Rel2}, \eqref{Rel3}.
\QED

\subsection{Solutions of the second Calogero--Painlev\'e systems from uncoupled solutions of PII}
\label{specsols}
We now go back to the matrix case and we assume that we are in the Kazhdan--Kostant--Sternberg setting, with

$$[\p,\q]=ig (\1 - v v^t)\in Mat_n(\mathbb{C}).$$

As described in the subsection \ref{SectionPII}, the eigenvalues $\{x_1,\dots, x_n\}$ of $\q$ satisfy 
\bea
\label{P2CalMos}
\ddot x_k = 2x_k^3 + t\,x_k + \theta + \sum_{ j \neq k}^n \frac {2 g^2}{(x_k-x_j)^3},\ \ \ k=1,\dots, n. 
\eea
Since the spectrum of $\1-v v^t$ is $\{1-n, 1\}$ with multiplicities $1$ and $n-1$, respectively, the operator  $Q= {\rm e}^{i\pi [\p,\q]}$ of the previous section  is $\pm \1$ for certain quantised values of the coupling parameter $g$.  
Specifically:
\bea
&&\hbox{If  $ig \in 2\Z$ and $n\in \N \Longrightarrow Q=\1$};\label{Q+1}
\\
&&\hbox{if $ig \in 2\Z+1$ and $n\in 2\N \Longrightarrow Q=-\1$.}
\label{Q-1}
\eea
In either situations the monodromy relations \eqref{qstokes} imply that the Stokes operators (matrices) $\A,\B,\C$ all commute (see also the Remark \ref{remclassic}) and {\it the exponents of formal monodromy are all integers (see \eqref{PhiForm2})}. In the generic situation, which we henceforth assume,  the Stokes operators are all simultaneously diagonalizable.  Indeed, without loss of  generality, we will assume that they are already diagonalized. In what follows we denote $ig = r$.
The goal of this section is to describe a method for finding solutions of the coupled Painlev\'e\ II Calogero--Moser system \eqref{P2CalMos}, starting from uncoupled ones and using discrete Schlesinger transformations \cite{JMU2}, \cite{BertolaCafasso5}.\\

Consider the uncoupled ($g=0$) Painlev\'e II system \eqref{PII}, and let $\Phi_0$ be the ``bare'' wave function of size $n\times n$ with $\A, \B, \C$ diagonal (semisimple) matrices satisfying the ``commutative'' relation 
\be
\A + \B + \C + \A \B\C =  2i\sin(\pi \theta) ,\  \ \ \A = {\rm diag}\left[a^{(j)}\right]_{j=1}^n,\ 
\B = {\rm diag}\left[b^{(j)}\right]_{j=1}^n,\ \C = {\rm diag}\left[c^{(j)}\right]_{j=1}^n
\label{commstokes}
\ee
It follows that the system \eqref{PII} is just a direct sum of $n$ copies of $2\times 2$ similar systems and
we denote by $\phi^{(j)}(z; a^{(j)}, b^{(j)}, c^{(j)}; \theta)$ the corresponding $2\times 2$ matrix solutions. 

We now set up the bare matrix with the goal of constructing the solution of the coupled problem with coupling constant $ig = r\in \Z$. Keeping in mind that when $r$ is odd (see \eqref{Q-1}) the monodromy relations \eqref{qstokes} reduce to the commutative case (but with $\theta\mapsto -\theta$), we construct the 
 $n\times n$ bare solution  in block form as 
\be
\Phi_0(z;\A,\B,\C;\theta) =  \le[
\begin{array}{c|c}
\ds {\rm diag}\left[\phi^{(j)}_{11}(z; a^{(j)}, b^{(j)}, c^{(j)}; (-)^r\theta) \right]_{j=1}^n &
\ds {\rm diag}\left[\phi^{(j)}_{12}(z; a^{(j)}, b^{(j)}, c^{(j)}; (-)^r\theta) \right]_{j=1}^n
\\[10pt]
\hline\\
\ds {\rm diag}\left[\phi^{(j)}_{21}(z; a^{(j)}, b^{(j)}, c^{(j)}; (-)^r\theta) \right]_{j=1}^n & 
\ds {\rm diag}\left[\phi^{(j)}_{22}(z; a^{(j)}, b^{(j)}, c^{(j)}; (-)^r\theta) \right]_{j=1}^n 
\end{array}
\ri]
\ee
so that each block is a diagonal matrix.
Because of \eqref{PhiForm2}, the matrix $\Phi_0$ has the following asymptotic behaviour at $z = \infty$ (cf. also \cite{ItsKapaevFokasBook})
\be
\Phi_0 \sim \le(\1 + \mathcal O(z^{-1}) \ri) z^{0} {\rm e}^{\frac i2 \le(\frac {z^3}3 + t z\ri) \wh \s_3},
\ee
where we emphasized that the exponents of formal monodromy at $\infty$ are all zero. The main result of this subsection is the following:

\begin{proposition}
	For any couple $(r = ig, n)$ as in \eqref{Q+1} or \eqref{Q-1} there exists a polynomial matrix $R(z)$ such that 
	$$\Phi(z) := R(z)\Phi_0(z)$$
	is a solution of the Lax system \eqref{PII} with $[\p,\q] = r(1 - v^Tv)$. Hence, in particular, the solution $\q$ of \eqref{P2CalMos} is given explicitely in terms of $\Phi_0$ using \eqref{PhiForm2}.
\end{proposition}
\noindent {\bf Proof:}
 Let  $\K$ be the diagonalizing matrix of $\1-v v^t$:
\be
\label{diaging}
\K^{-1} (\1-v v^t) \K = {\rm diag}(1-n, 1,\dots, 1),
\ee 
and consider  
$$\wt \Phi_0(z) := (\K\ot \1) \Phi_0(z;\A,\B,\C;\theta)(\K\ot\1)^{-1},$$ 
which is still a solution of \eqref{PII} with all the exponents of the formal monodromy at infinity equal to $0$ (but non--diagonal Stokes parameters). Thanks of the classical work of Jimbo Miwa and Ueno \cite{JMU2} (see also \cite{BertolaCafasso5, BertolaIsoTau}) it is known that there exists a polynomial matrix $\wt R(z)$  of degree $r n$ such that the matrix
\be
\wt \Phi(z;\A,\B,\C;\theta):= \wt R(z) \wt \Phi_0(z;\A,\B,\C;\theta)
\ee 
has exponents of formal monodromy at $\infty$  equal to $r{\rm diag}(1-n, 1,\dots, 1)\ot \1$. In other words, at infinity we have that
\be
\wt \Phi(z;\A,\B,\C;\theta)
=
  \le(\1 + \mathcal O(z^{-1}) \ri) z^{r {\rm diag}(1-n, 1,\dots, 1)\ot \1} {\rm e}^{\frac i2 \le(\frac {z^3}3 + t z\ri) \wh \s_3}.
\ee
Indeed the construction of $\wt R(z)$ amounts to a large but finite set of linear equations; it can be constructed iteratively in terms of ``elementary'' Schlesinger transformations that shift by $+1$ and $-1$ exactly two exponents of formal monodromy. The resulting formulas are completely explicit (in terms of the coefficients of the $z^{-1}$--expansion of $\Phi_0$) but extremely large, even for the simplest case $n=2, r=\pm 1$. Finally, in order to replace $z^{r\diag(1-n,1,\ldots,1)}$ in the equation above with $z^{ig(\1 - v^Tv)}$, we simply define
\be
\Phi (z;\A,\B,\C;\theta) :=   (\K\ot \1)^{-1} \wt \Phi(z;\A,\B,\C;\theta) (\K\ot \1).
\ee
In particular, we have that the matrix $R(z)$ to be found is equal to
\be
	R(z) := (\K\ot\1)^{-1}\wt R(z) (\K\ot\1).
\ee
\QED


As a Hamiltonian system, the particle--particle interaction   in \eqref{P2CalMos} is {\it attractive} and hence from a dynamical point of view it would be not clear whether there  exist  global solutions $\vec x(t)$. Indeed the Hamiltonian is not bounded below.  However, we have constructed explicit solutions starting from solutions of the single--particle equations; these solutions $\q(t)$ necessarily  have the Painlev\'e\ property (they are ultimately very complicated rational expressions in the single--particle solutions and their derivatives) and hence, even if they may have poles for some real values of $t$, they are globally defined; of course this does not prevent the eigenvalues to collide for some values of $t$.

\paragraph{Acknowledgements.} 
We thank P. Boalch for pointing out the relations between this work and his article \cite{Boalch}, and G. Rembado for some interesting discussions on the deformation quantization of simply laced isomonodromy systems.\\
The three authors acknowledge the support of the project IPaDEGAN (H2020-MSCA-RISE-2017), grant number 778010.
The research of M.B  was supported in part by the Natural Sciences and Engineering Research Council of Canada grant
RGPIN-2016-06660 and by the FQRNT grant "Applications des syst\`emes int\'egrables aux surfaces de Riemann et aux espaces de modules".
The research of M.C. and V.R. was partially supported by a project ``Nouvelle \'equipe'' funded by the region Pays de la Loire.
V. R. acknowledges the support of the Russian Foundation for Basic Research under the grants RFBR 18-01-00461 and 16-51-53034-GFEN.
M.C. and V.R. thank the Centre ``Henri
Lebesgue'' ANR-11-LABX-0020-01 for creating an attractive mathematical environment. They also thank the International School of Advanced Studies (SISSA) in Trieste for the hospitality during part of the preparation of this work.


\def\cprime{$'$}


\begin{thebibliography}{10}

\bibitem{BertolaIsoTau}
M.~Bertola.
\newblock The dependence on the monodromy data of the isomonodromic tau
  function.
\newblock {\em Comm. Math. Phys.}, 294(2):539--579, 2010.

\bibitem{BertolaCafasso2}
M.~Bertola and M.~Cafasso.
\newblock {Fredholm determinants and pole-free solutions to the noncommutative
  Painleve' II equation}.
\newblock {\em Comm. Math. Phys.}, 309(3):793--833, 2012.

\bibitem{BertolaCafasso5}
M.~Bertola and M.~Cafasso.
\newblock Darboux transformations and random point processes.
\newblock {\em International Mathematics Research Notices}, 15:6211 -- 6266,
  2015.

\bibitem{BloemViragI}
A. Bloemendal and B. Vir{\'a}g.
\newblock {Limits of spiked random matrices I}.
\newblock {\em Probability Theory and Related Fields}, 156, 2013.


\bibitem{BEMN}
G. Borot, B. Eynard, S. N. Majumdar and C. Nadal.
\newblock{Large deviations of the maximal eigenvalue of random matrices}.
\newblock{\em Journal of Statistical Mechanics: Theory and Experiment}, P11024, 2011.


\bibitem{FN}
H. Flaschka and A.~C. Newell.
\newblock Monodromy- and spectrum-preserving deformations. {I}.
\newblock {\em Comm. Math. Phys.}, 76(1):65--116, 1980.

\bibitem{ItsKapaevFokasBook}
A.~S. Fokas, A.~R. Its, A.~A. Kapaev, and V.~Yu.
  Novokshenov.
\newblock {\em Painlev{\'e} transcendents}, volume 128 of {\em Mathematical
  Surveys and Monographs}.
\newblock American Mathematical Society, Providence, RI, 2006.
\newblock The Riemann-Hilbert approach.

\bibitem{Ino}
V.~I. Inozemtsev.
\newblock {Lax representation with spectral parameter on a torus for integrable
  particle systems}.
\newblock {\em Lett. Math. Phys.}, 17:11--17, 1989.

\bibitem{ItoTerwilliger}
T. Ito and P. Terwilliger.
\newblock Double affine {H}ecke algebras of rank 1 and the {$\mathbb
  Z_3$}-symmetric {A}skey-{W}ilson relations.
\newblock {\em SIGMA Symmetry Integrability Geom. Methods Appl.}, 6:Paper 065,
  9, 2010.

\bibitem{JMU2}
M.~Jimbo and T.~Miwa.
\newblock Monodromy preserving deformation of linear ordinary differential
  equations with rational coefficients. {II}.
\newblock {\em Phys. D}, 2(3):407--448, 1981.
\bibitem{JMU1}
M.~Jimbo T.~Miwa  and K.~Ueno.
\newblock Monodromy preserving deformation of linear ordinary differential
  equations with rational coefficients. {I}. {G}eneral theory and {$\tau
  $}-function.
\newblock {\em Phys. D}, 2(2):306--352, 1981.


\bibitem{cubicpencils}
K.~Kajiwara, T.~Masuda, M.~Noumi, Y.~Ohta, and Y.~Yamada.
\newblock Cubic pencils and {P}ainlev{\'e} {H}amiltonians.
\newblock {\em Funkcial. Ekvac.}, 48(1):147--160, 2005.

\bibitem{Kawa1}
H.~Kawakami.
\newblock {Matrix Painlev\'e Systems}.
\newblock {\em J. Math. Phys.}, 56, 2015.

\bibitem{Kawa2}
H.~Kawakami.
\newblock {Four-dimensional Painlev{\'e}-type equations associated with
  ramified linear equations I: Matrix Painlev\'e systems}.
\newblock {\em arXiv: 1608.03927}, 2016.

\bibitem{KKS}
D.~Kazhdan B.~Konstant  and S.~Sternberg.
\newblock {Hamiltonian group actions and dynamical systems of Calogero type}.
\newblock {\em Communications on Pure and Applied Mathematics}, 1978.

\bibitem{KricheverCM}
I.~Krichever.
\newblock {On rational solutions of the Kadomtsev-Petviashvili equation and
  integrable systems of $N$ particles on the line.}
\newblock {\em Funct. Anal. Appl.}, 12:1, 1978.

\bibitem{LeOl}
A.~M. Levin and M.~A. Olshanetsky.
\newblock {Painlev\'e-Calogero Correspondence}.
\newblock In {\em Calogero-Moser-Sutherland Models}, CRM Series in Mathematical
  Physics. CRM, 2000.

\bibitem{ManinPainleve}
Yu.~I. Manin.
\newblock Rational curves, elliptic curves, and the {P}ainlev{\'e} equation.
\newblock In {\em Surveys in modern mathematics}, volume 321 of {\em London
  Math. Soc. Lecture Note Ser.}, pages 24--33. Cambridge Univ. Press,
  Cambridge, 2005.

\bibitem{MazRou} M.~Mazzocco and
V.~Rubtsov.
\newblock {Confluence on the Painlev{\'e} Monodromy Manifolds, their Poisson
  Structure and Quantisation}.
\newblock {\em arXiv: 1212.6723}, 2012.


\bibitem{Mazzocco13}
M. Mazzocco.
\newblock Confluences of the {P}ainlev{\'e} equations, {C}herednik algebras and
  {$q$}-{A}skey scheme.
\newblock {\em Nonlinearity}, 29(9):2565--2608, 2016.

\bibitem{OkaHam}
K.~Okamoto.
\newblock {Polynomial Hamiltonians associated with Painlev{\'e} equations, I.}
\newblock {\em Proceedings of the Japan Academy, Series A, Mathematical
  Sciences.}, 56, 1980.

\bibitem{OkamotoSurfaces}
K. Okamoto.
\newblock Sur les feuilletages associ{\'e}s aux {\'e}quations du second ordre
  {\`a} points critiques fixes de {P}. {P}ainlev{\'e}.
\newblock {\em Japan. J. Math. (N.S.)}, 5(1):1--79, 1979.

\bibitem{Painleve}
P.~Painlev\'e.
\newblock {Sur les \'equations diff\'erentielles du second ordre \`a points
  critiques fix\'es}.
\newblock {\em C.R. Acad. Sci (Paris)}, 143:1111--1117, 1906.

\bibitem{Rembado}
G. Rembado.
\newblock {Simply-laced quantum connections generalising KZ}.
\newblock{\em arXiv: 1704.08616}, 2017.

\bibitem{Reshetikhin}
N. Reshetikhin. 
\newblock{The Knizhnik-Zamolodchikov system as a deformation of the isomonodromy problem}. 
\newblock{\em Lett. Math. Phys.}, 26(3):167--177, 1992.

\bibitem{RetakhRubtsov}
V. Retakh and V. Rubtsov.
\newblock {Noncommutative Toda Chains, Hankel Quasideterminants and Painlev\'e
  II Equation}.
\newblock {\em Journal of Physics A: Mathematical and General}, 43(50), 2010.

\bibitem{RumanovqPII}
I.~Rumanov.
\newblock {Classical integrability for beta-ensembles and general Fokker-Planck
  equations.}
\newblock {\em J. Math. Phys.}, 56, 2015.

\bibitem{ShiodaTakano}
T. Shioda and K. Takano.
\newblock On some {H}amiltonian structures of {P}ainlev{\'e} systems. {I}.
\newblock {\em Funkcial. Ekvac.}, 40(2):271--291, 1997.

\bibitem{TakasakiPaCa}
K. Takasaki.
\newblock Painlev{\'e}-{C}alogero correspondence revisited.
\newblock {\em J. Math. Phys.}, 42(3):1443--1473, 2001.

\bibitem{TracyWidomLevel}
C.~A. Tracy and H.~Widom.
\newblock Level-spacing distributions and the {A}iry kernel.
\newblock {\em Comm. Math. Phys.}, 159(1):151--174, 1994.

\bibitem{SaitovanderPut}
M. van~der Put and M.-H. Saito.
\newblock Moduli spaces for linear differential equations and the
  {P}ainlev{\'e} equations.
\newblock {\em Ann. Inst. Fourier (Grenoble)}, 59(7):2611--2667, 2009.

\bibitem{Wasow}
W. Wasow.
\newblock {\em Asymptotic expansions for ordinary differential equations}.
\newblock Dover Publications Inc., New York, 1987.
\newblock Reprint of the 1976 edition.

\bibitem{WilsonAdelic}
G. Wilson.
\newblock Collisions of {C}alogero-{M}oser particles and an adelic
  {G}rassmannian.
\newblock {\em Invent. Math.}, 133(1):1--41, 1998.
\newblock With an appendix by I. G. Macdonald.

\bibitem{QPC1ZZ}
A.~Zabrodin and A.~Zotov.
\newblock Quantum {P}ainlev{\'e}-{C}alogero correspondence.
\newblock {\em J. Math. Phys.}, 53(7):073507, 19, 2012.

\bibitem{QPC2ZZ}
A.~Zabrodin and A.~Zotov.
\newblock Quantum {P}ainlev{\'e}-{C}alogero correspondence for {P}ainlev{\'e}
  {VI}.
\newblock {\em J. Math. Phys.}, 53(7):073508, 19, 2012.

\end{thebibliography}
\end{document}